\documentclass[twocolumn]{aastex6}

\pdfoutput=1

\usepackage{graphicx}
\slugcomment{Accepted for publication in ApJ}

\shorttitle{Star Formation History of the Carina Galaxy.}
\shortauthors{Santana et al.}
\begin{document}

\title{A MegaCam Survey of Outer Halo Satellites. VI: The Spatially Resolved Star Formation 
History of the Carina Dwarf Spheroidal Galaxy\altaffilmark{1}} 
\author{
Felipe\ A.\ Santana$^{2}$,
Ricardo\ R.\ Mu\~noz$^{2}$,
T.\ J. L. de Boer$^{3}$,
Joshua\ D.\ Simon$^{4}$,
Marla\ Geha$^{5}$,
Patrick\ C\^ot\'e$^{6}$,
Andr\'es\ E.\ Guzm\'an$^{2}$,
Peter\ Stetson$^{6}$, and
S.\ G.\ Djorgovski$^{7}$
}

\altaffiltext{1}{Based on observations obtained with the MegaCam imager on the
Magellan~II-Clay telescope at  Las Campanas Observatory in the Atacama Region, Chile.
This telescope is operated by a consortium consisting of the Carnegie Institution of 
Washington, Harvard University, MIT, the University of Michigan, and the University of 
Arizona.}

\affil{%
 (2) Departamento de Astronom\'ia, Universidad de Chile, Camino El 
Observatorio 1515, Las Condes, Santiago, Chile (fsantana@das.uchile.cl, 
rmunoz@das.uchile.cl)
\\
 (3) Institute of Astronomy, University of Cambridge, Madingley Road, Cambridge CB3 0HA
UK
\\
 (4) Observatories of the Carnegie Institution of Washington, 813 Santa 
Barbara St., Pasadena, CA 91101, USA
\\
 (5) Astronomy Department, Yale University, New Haven, CT 06520, USA
\\
 (6) National Research Council of Canada, Herzberg Astronomy and Astrophysics Program, 
Victoria, BC, V9E 2E7, Canada
\\
 (7) Astronomy Department, California Institute of Technology, Pasadena, CA, 
91125, USA
}

\begin{abstract}

We present the spatially resolved star formation history (SFH) of the Carina dwarf spheroidal 
galaxy, obtained from deep, wide-field g,r imaging and a metallicity distribution from the 
literature.
Our photometry covers $\sim2$\,deg$^2$, reaching up to 
$\sim10$ times the half-light radius of Carina with a completeness higher than 
$50\%$ at $g\sim24.5$, more than one magnitude fainter than the oldest 
turnoff. 
This is the first time a combination of depth and coverage of this quality has been used to
derive the SFH of Carina, enabling us to trace its different populations with unprecedented 
accuracy.
We find that Carina's SFH consists of two episodes well separated by a star formation 
temporal gap.
These episodes occurred at old ($>10$\,Gyr) and intermediate ($2$--$8$\,Gyr) ages. 
Our measurements show that the old episode comprises the 
majority of the population, accounting for $54\pm5\%$ of the stellar mass within $1.3$
times the King tidal radius, while the total stellar mass derived for Carina is
$1.60\pm0.09\times 10^{6} \, M_{\rm{\odot}}$, and the stellar mass-to-light ratio
$1.8\pm0.2$.
The SFH derived is consistent with no recent star formation which hints that the observed
blue plume is due to blue stragglers.
We conclude that the SFH of Carina evolved independently of the tidal field of the Milky
Way, since the frequency and duration of its star formation events do not correlate with its 
orbital parameters.
This result is supported by the age/metallicity relation observed in Carina, and
the gradients calculated indicating that outer regions are older and more metal poor.

\end{abstract}

\keywords{Local Group ; galaxies: stellar content ; galaxies: evolution ; galaxies: dwarf ;
Hertzsprung-Russell and C-M diagrams}

\section{Introduction}

Dwarf galaxies are crucial for understanding galaxy assembly and evolution.
They are some of the oldest systems in the Universe, and inhabit the most numerous
type of dark matter halos in the framework of a Lambda-CDM Universe 
\citep[e.g.,][]{kauffmann93a}.
These systems gave origin to larger galaxies like the Milky Way in 
the early Universe via hierarchical merging \citep[e.g.,][]{unavane96}.
The dwarf galaxies in the Local Group are particularly interesting,
since their proximity allows us to resolve them into individual stars.
Thus, it is not surprising that these galaxies have been studied in more detail than any
other galaxies \citep{tolstoy09,mcconnachie12}.

Carina is a Local Group dwarf spheroidal (dSph) galaxy located at about $104$\,kpc from 
the Sun \citep{karczmarek15}, with a half-light radius of
$250\pm39\,$pc \citep{irwin95a},
a dynamical mass within the half-light radius of
$M_{\rm dyn}(<r_{\rm half})=3.4\pm1.4\times10^{6}\,$M$_{\odot}$ \citep{walker09a},
and an absolute magnitude of M$_{V}=-9.3$ \citep{mateo98a}.
This galaxy is especially important as a constraint on Galactic evolution since it is one of
the few Local Group galaxies showing an episodic star formation history (SFH)
\citep[e.g.,][]{weisz14b}.
Furthermore, Carina is the only galaxy where the episodic star formation history
translates into two clearly distinct sub-giant branches \citep[see for example][]
{bono10,deBoer14}. 
These star formation episodes may be either related to interactions with the Milky Way or 
with internal evolution of its gas and stars.

The SFH of local systems like Carina have been studied mainly through the analysis
of their color-magnitude diagrams (CMDs). In recent years this has been done using
the synthetic CMD method \citep[e.g.,][]{deBoer14,weisz14b}.
This technique consists of deriving the history of a stellar system by 
creating different combinations of synthetic single stellar populations and comparing their
properties to the ones of the stars observed.
Carina was first thought to be a purely intermediate-age
population galaxy, but then RR-Lyraes were discovered in this system, indicating that
an old ($>10$\,Gyr) stellar population was also present in this galaxy \citep{saha86}. 
Multiple main-sequence turnoffs confirmed that Carina had an episodic SFH
\citep{smecker-hane96,hurley-keller98}, which means that there are clearly
distinguishable episodes of star formation activity separated by
episodes where practically no stars were formed.
Another key feature of Carina's CMD is its narrow red giant branch (RGB), which was at first 
interpreted as the result of a low metallicity spread \citep[see][and references therein]
{rizzi03}.
In that work, the authors measured a color spread of the RGB of
$\sigma_{\rm{V-I}}=0.021\pm0.005$ and derived a metallicity of [Fe/H]=$-1.91$ with a
 spread of $0.12$\,dex, in agreement with early spectroscopic studies of upper RGB stars in
Carina \citep[e.g.,][]{armandroff91}.
More recent spectroscopic observations have 
detected a  much larger metallicity spread in this galaxy \citep{helmi06a,koch06}.
The latter study measured a mean metallicity of [Fe/H]$\sim-1.4$ and
a spread of $0.92$\,dex.
More recently, \citet{deBoer14} used Koch's metallicity distribution function (MDF) along
with CMD information from their deep photometry and derived a self-consistent, complex
SFH for Carina, indicating a strong age/metallicity degeneracy. This degeneracy implies
that some properties of a given stellar population, like its MDF or its RGB color and width,
are equivalent to the ones of a population that is older and more metal poor.
Confirmation or refutation of these results would shed important light on the origin of 
Carina's photometric and spectroscopic features.

From all these previous works there is a general agreement that Carina has a well separated,
episodic SFH consisting of at least two episodes producing old an intermediate-age 
populations.
In addition, several studies \citep{hurley-keller98,mateo98b,monelli03} claimed
a third episode in Carina, consisting of young  ($<1$\,Gyr) stars.
However, the exact age and duration of all these potential episodes are still uncertain.

Another important feature found in Carina is the metallicity and age radial gradients of its stellar
populations.
In Carina's outer regions, the relative prevalence of older and more 
metal-poor stars increases \citep{munoz06b,battaglia12,mcMonigal14,deBoer14}.
One of these contributions, \citet{munoz06b} presented several observations that 
were indicative of tidal effects on Carina, a plausible explanation for these population 
gradients.
These pieces of evidence include a component in the density profile that extends well
beyond the nominal King limiting radius, a distribution of outer stars that lie preferentially
along the major axis and a velocity dispersion profile that rises well past the limiting radius. 
These results were then 
supported 
by the work of \citet{munoz08a} and \citet{battaglia12}. 
It is worth mentioning that the proper motion study of Carina by \citet{piatek03}
results in an orbit that is consistent with the tidal scenario proposed by \citet{munoz08a}.
In their study, \citet{piatek03} claimed that this dSph is currently at apocenter, has an orbital period
close to $2$\,Gyr and its last close passage from the Milky Way occurred $\sim0.7\,$Gyr ago.
Tidal effects are interesting in the context of the work by \citet{piatek03}
and \citet{pasetto11} who tested if the star formation episodes of Carina 
could be explained as the result of close encounters with the Milky Way.
These encounters can promote star formation for example by removing angular momentum
from the gas and driving it to the central regions \citep[e.g.,][]{larson02}, or by compression 
produced by tidal gravitational shocks \citep[e.g.,][]{pasetto11}.
They placed constraints on Carina's orbit, but have achieved limited 
success in explaining Carina's SFH as a result of tidal shocks or ram pressure.

Another proposed origin for the properties of Carina's SFH is internal evolution.
For example, gas depletion or radial migration \citep{el-badry15} might produce the 
positive age radial gradient (average stellar population age increasing with radius) and the 
negative metallicity gradient (average metal content decreasing with radius). 
Additionally, gas heating \citep[see 
for example][]{revaz09} might explain the temporal gap in star formation.

In summary, 
we do not have at present a complete
scenario explaining the evolution of Carina that
is consistent with its SFH, chemical enrichment, orbital information and gas dynamics.
In this work we use deep/wide photometry along with public metallicities to derive
the SFH of Carina.
By making use of the Talos routine \citep{deBoer12} we take into consideration all the 
information in the CMD (and not just some key fingerprints) along with the MDF to derive
the SFH in a consistent way.
The high quality of the photometry along with the spatial extent of the observations
(two square degrees), enable us to make three independent measures of the Carina's SFH
within different concentric regions.
In this way, we can quantify the dependence of the SFH on the distance from the galaxy 
center by looking at radial gradients.

In Section~2 we present the spectroscopic and photometric data. Section~3 describes the
method for deriving the SFH along with the input files used. Section~4 presents the main 
results for the SFH which are discussed in Section~5.

\section{Data} 

The observations used in this article are part of a larger survey to obtain deep/wide photometry
of all the Milky Way satellites in the outer halo (R.~R.~ Mu\~noz et al., in preparation).
The main structural parameters of Carina used in this study were derived in that same work 
by fitting different density profiles to the data and using a Maximum Likelihood method. 
From this work we took the King tidal radius of 
Carina ($r_{\rm{tidal}}=31{\farcm}0\pm0.3$), which was estimated as the limiting radius of 
the King (1962) profile. The other structural parameters used from this study are the central
coordinates of Carina ($\alpha_{\rm{0}} = 06\fh 41\fm 01\fs70 $, $\delta_{\rm{0}}=
-50^{\circ}57{\farcm}58{\farcs}0$), the ellipticity ($\epsilon=0.35\pm0.01$), and the
position angle ($\theta=61\pm1$).

\begin{figure*}
\includegraphics[width=\textwidth]{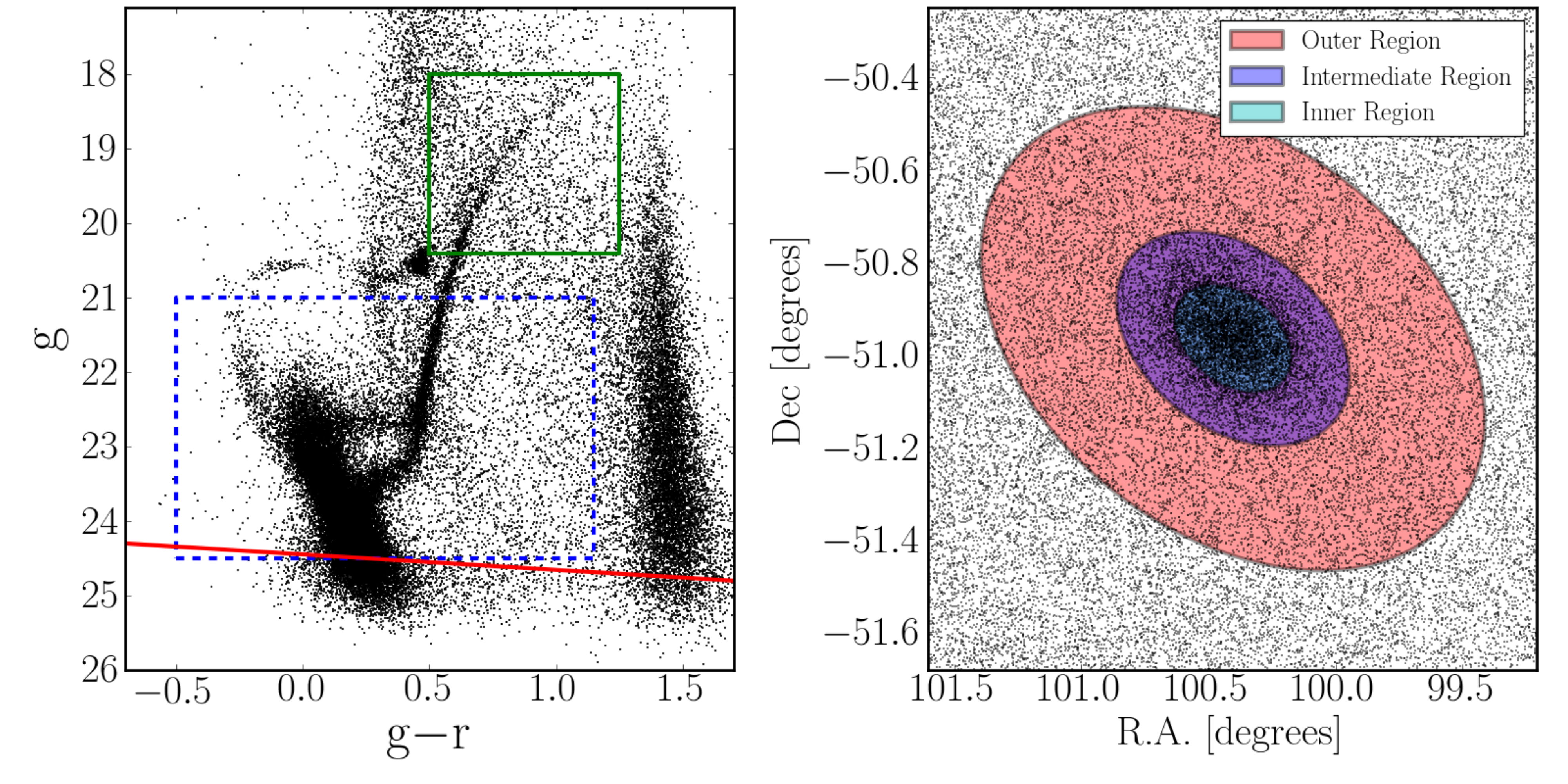}
\caption{ Stars in Carina field. Left: g v.s. g-r dereddened CMD including 
all stars within $1.3\times r_{\rm{tidal}}$ of Carina, where stars have been selected 
according to their photometric errors and a chi/sharp criteria.
Red solid line shows the 50\% completeness level. The dashed blue box shows the CMD 
region used to compare the colors and magnitudes of the data with the models 
(photometric CMD region) to derive the SFH, while the green solid box shows the CMD 
region used to compare the metallicities of the data and the different models (spectroscopic
CMD region) for the SFH derivation (see details in Section~3.2).
Right: Star map of Carina, showing the 
different regions where the SFH was determined. Inner region: $0<r/r_{\rm{tidal}}<0.3$.
Intermediate region: $0.3<r/r_{\rm{tidal}}<0.6$. Outer region: $0.6<r/r_{\rm{tidal}}<1.3$. }
\label{cmd_and_starmap}
\end{figure*}

\begin{figure*}
\includegraphics[width=\textwidth]{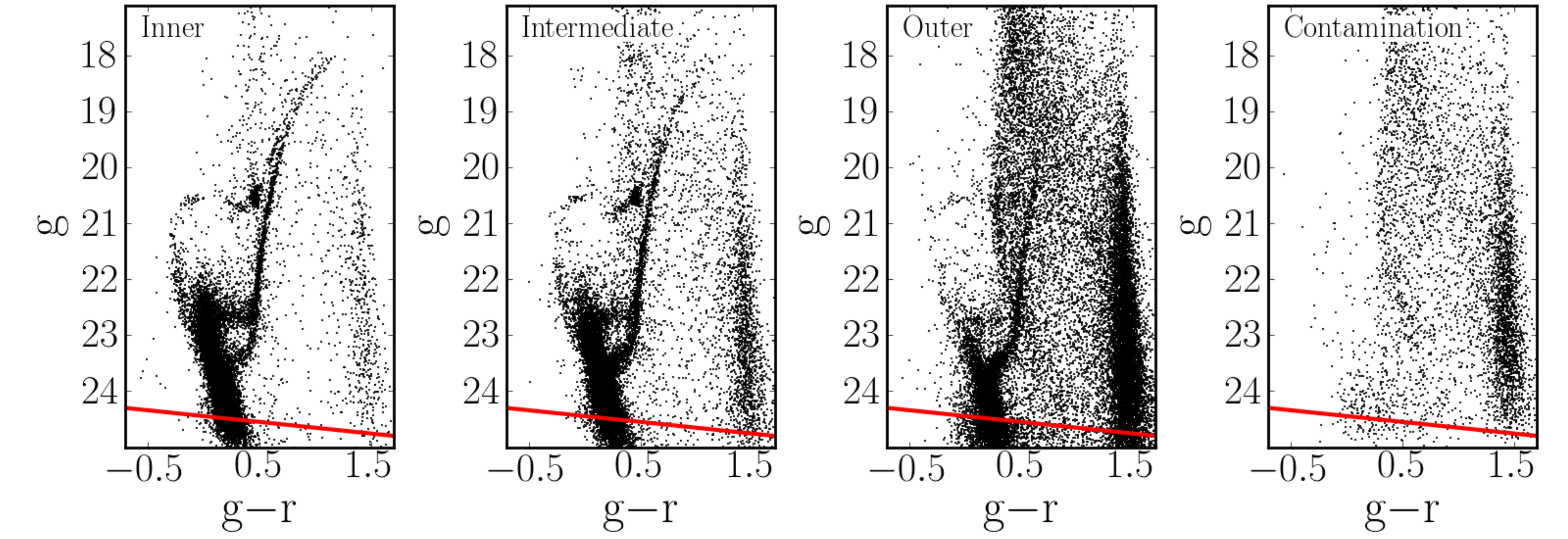}
\caption{ Dereddened CMDs of the different regions defined for Carina.
The stars in these plots have been selected according to their photometric errors and a
chi/sharp criteria. Red solid line shows the 50\% completeness level. 
From left to right we show the Inner region ($0<r/r_{\rm{tidal}}<0.3$),
Intermediate region ($0.3<r/r_{\rm{tidal}}<0.6$), Outer region ($0.6<r/r_{\rm{tidal}}<1.3$) 
, and the region used to calculate the background/foreground contamination
($2.0<r/r_{\rm{tidal}}<3.0$).
}
\label{cmds_diff_regions}
\end{figure*}

\subsection{Photometry}

Images for Carina were obtained using the MegaCam imager 
on the Magellan~II-Clay telescope at  Las Campanas 
Observatory in the Atacama Region, Chile.
A total of $16$ fields in a $4 \times 4$ configuration were observed, centered on Carina, 
achieving a total area of $\sim2$\,deg$^2$.
This translates into a full coverage within $1.3$ times the tidal radius of Carina and partial
coverage from $1.3$ to $3.0\times r_{\rm{tidal}}$.
For each field observed, we took five dithered exposures in the
Clay $g-$ and $r-$ bands, using a standard dithering pattern from the 
MegaCam operation options in order to cover the gaps between the chips.
The exposure times were $90$\,s for the $g-$ images and $180$\,s for the $r-$ images.

The images were first processed to correct for bias levels, variations in the pixel-to-pixel
sensitivity, and bad pixels.
Point spread function photometry was then carried out simultaneously in the overlapping 
images by using the packages DAOPHOT/Allstar/ALLFRAME \citep{stetson94a}, as
described in \citet{munoz10a}.
From all the sources obtained with our photometry, we selected the ones that are most
likely to correspond to stars. To make this distinction, we used the 
DAOPHOT parameters {\it Chi} and {\it sharpness}, indicating the significance of the point
spread function fitting and the thickness of the brightness profile, respectively.
The sources selected as stars were the ones with magnitude uncertainties smaller than
$0.1$ in both $g-$ and $r-$ bands, and satisfying $Chi < 5$, and $-0.4<sharp<0.4$.
Finally, we calibrated our photometry by transforming our Clay $g-$ and $r-$ instrumental 
magnitudes into the Sloan Digital Sky Survey \citep{york00a} $g-$ and $r-$ 
magnitudes.
For this, we used the photometry from the globular clusters and dwarf galaxies from our
photometric catalogue of Milky Way satellites that overlap the Sloan footprint.
By comparing the magnitudes of the stars in these systems with those in Sloan, we
derived the zero-points, color and extinction terms that were then applied to calibrate
the magnitudes of the stars in our Carina data.
Figure~\ref{cmd_and_starmap} shows the spatial coverage along with
the final calibrated CMD of all the sources selected as stars.

\subsection{Spectroscopy}

We used archival, medium resolution ($R \sim 6500$) Ca II triplet spectroscopy from 
\citet{koch06}.
The targets were chosen from the ESO imaging
study \citep{nonino99} sample, according to their photometry and astrometry.
From that sample, the authors selected the stars across the full width of the RGB (0.2 
magnitudes in $B-V$), from the RGB tip down to 3 magnitudes below the RGB tip.
By choosing a wide color range, they ensured that potentially extremely metal-poor
or metal-rich giants were included.
The complete spectroscopic sample of \citet{koch06} included $1257$ stars in the region
centered on Carina, covering approximately $1$\,deg$^{2}$, and reaching in all cases
a signal-to-noise ratio of at least $20$.
Then, the members of the Carina galaxy were chosen as the sources with radial velocities
consistent with the systematic heliocentric radial velocity of Carina of $\sim220$\,km 
s$^{-1}$.
The final catalogue of kinematic members of Carina from \citet{koch06} contains 
$437$ sources.
From this sample, $430$ were matched to stars in our photometric data by selecting sources 
that were closer than $1$\,arcsecond from a star belonging to our 
sample\footnote{Given that the typical distances from stars in our photometric sample are
at the order of tens of arcseconds even in the highest density regions, this matching criteria
is extremely unlikely to produce matches between observations that do not correspond to
the same star.}.
Both the magnitudes and the positions of the spectroscopic sample were taken from our
photometric catalogues to make them directly comparable to the photometric sample.
For determining the iron abundances, \citet{koch06} translated the
equivalent widths of Ca II using the calibrations of \citet{zinn84} and \citet{carretta97}.
However, in this study we used the calibration of \citet{starkenburg10} to obtain the
[Fe/H] values, mainly to avoid saturation at low equivalent widths.
Our final sample comprises measurements out to $1.0\times r_{\rm{tidal}}$
of Carina, and spans a range in metallicity of $-3.8 < $[Fe/H]$ < 0.0 $.

\section{SFH derivation Method}

To derive the SFH of Carina we used the Talos routine \citep{deBoer12}. In this section
we describe its main features, how it was implemented for the case of our Carina
data, and the construction of the necessary input files.

\subsection{Talos description}

The Talos routine is a set of Fortran scripts developed to derive the SFH
of an object based on its CMD and its MDF. It uses a set of isochrones chosen by the user, 
which at the moment includes the options Teramo \citep{pietrinferni04}, Dartmouth 
\citep{dotter08} and Yonsei-Yale \citep{spada13}. With the isochrones chosen, it
creates synthetic populations that are then contrasted to observations.
To populate the synthetic models, Talos uses a Kroupa IMF \citep{kroupa01}, and takes into
account various properties of the input data to make the models directly comparable to the
observations.
These properties include distance, reddening, binary fraction, photometric errors and 
completeness fractions.
Then, the combination of synthetic populations that best matches the observed
decontaminated CMD and MDF is selected as the SFH of the object.
The comparison between synthetic and observed CMDs is done using Hess diagrams, in
which each point represents the number of stars at a specific color/magnitude bin.
Each time a combination of synthetic populations is created, the routine calculates the
differences between the star counts in the model and the observations at each 
color/magnitude bin; along with the difference at each metallicity bin from the MDF.
With these differences, Talos calculates a poissonian chi-squared, whose minimization
gives the resulting SFH  (i.e., the model that best reproduces simultaneously the observed 
CMD and MDF).

\subsection{General Setup}

\begin{figure*}
\includegraphics[width=\textwidth]{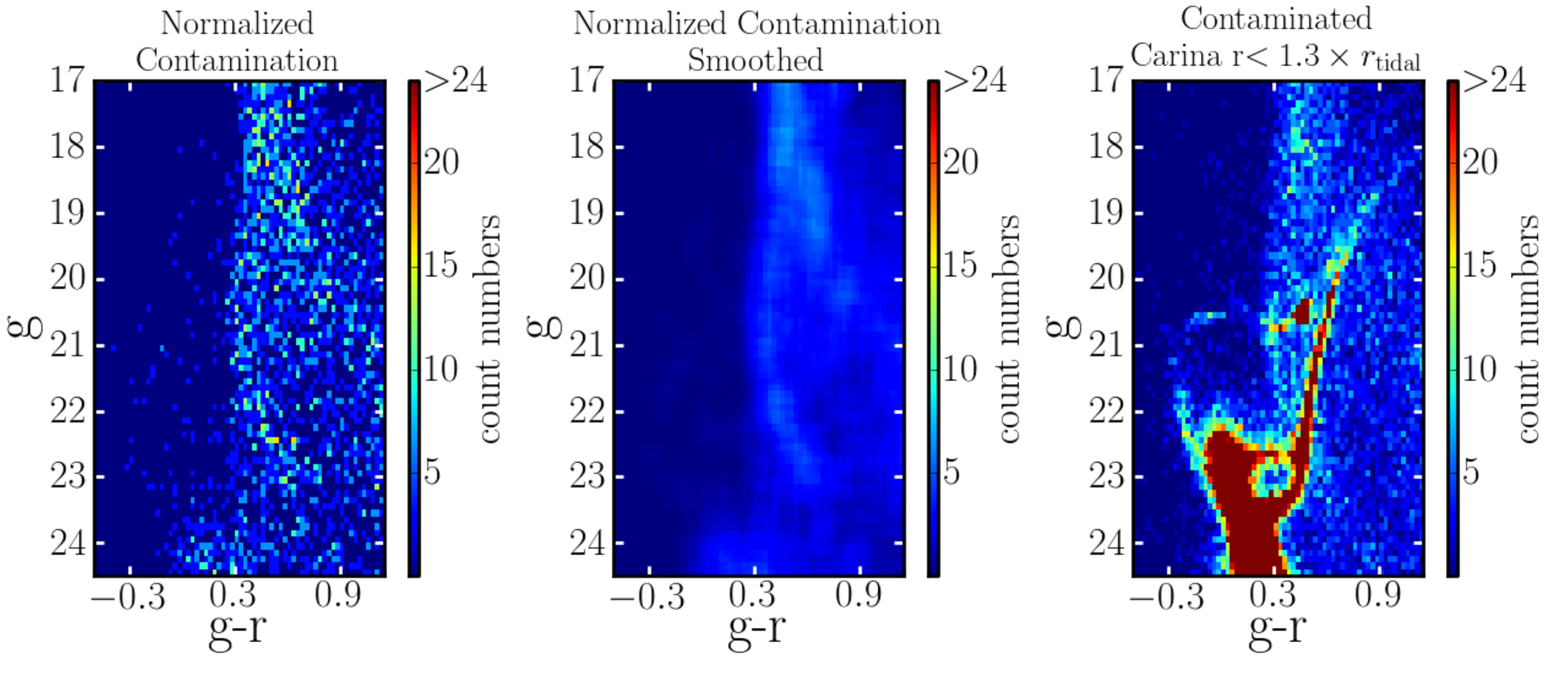}
\caption{Contamination smoothing process represented by Hess diagrams in which the 
different colors indicates the number of stars in each color-magnitude bin.
Left: Hess diagram of contamination region ($r>2.0\times r_{\rm{tidal}}$), counts have been
normalized multiplying the area of the contamination region over the area of the complete
Carina region ($r<1.3\times r_{\rm{tidal}}$).
Middle: contamination region normalized Hess diagram smoothed using a $6\times6$
gaussian Kernel convolution.
Right: Hess diagram of the complete Carina region without subtracting the contamination.
Shot noise is significantly reduced in the  smoothed diagram while still preserving the most 
important features that are also seen in the Hess diagram of Carina.}
\label{contamination}
\end{figure*}

\begin{figure}
\includegraphics[width=\columnwidth]{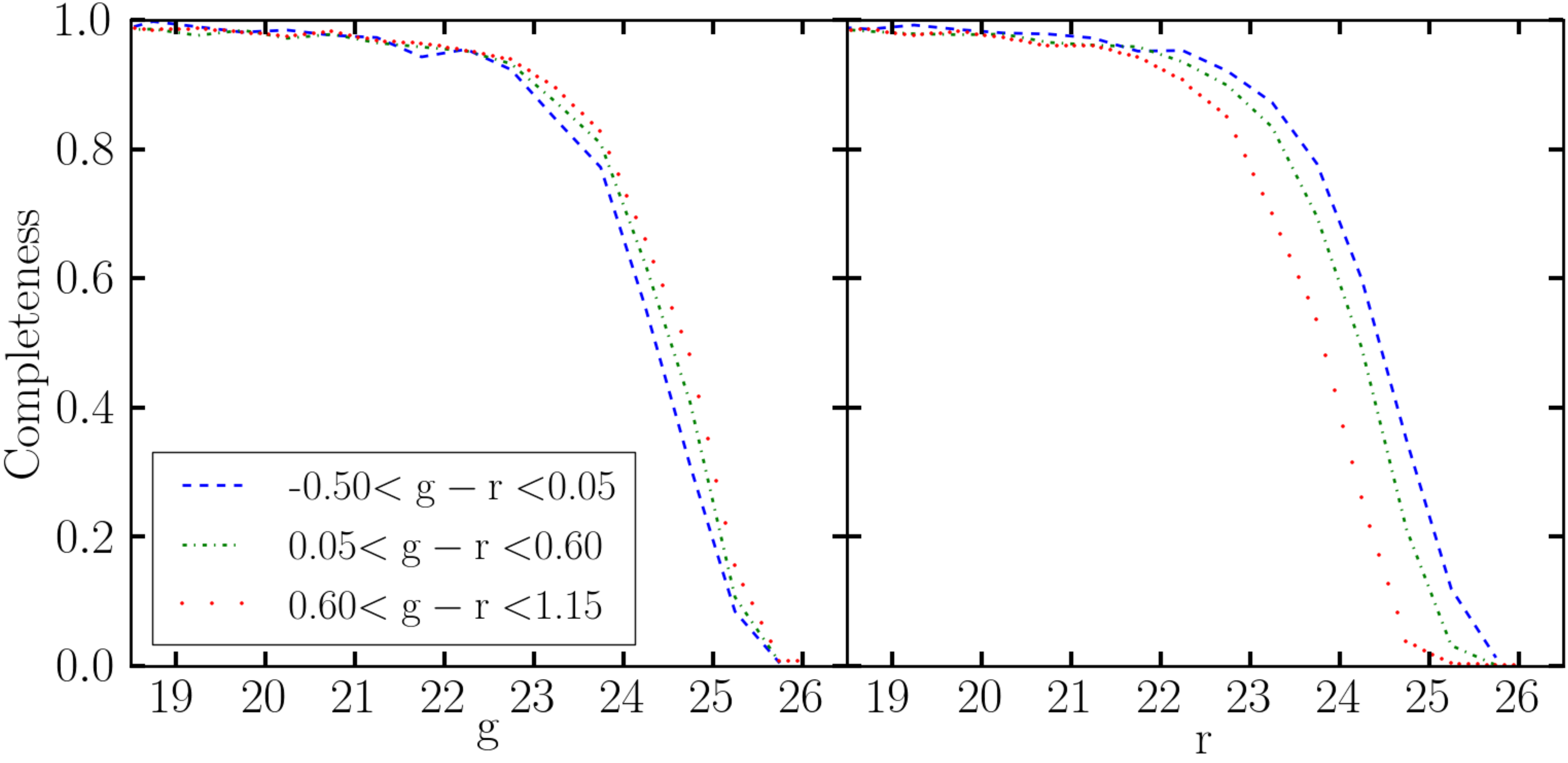}
\caption{ Completeness v.s. magnitude. Fraction of stars recovered by the artificial 
photometry as a function of input $g-$ (left) and $r-$ (right) magnitudes.
The different lines show stars at different input colors.}
\label{completeness_vs_mag}
\end{figure}

\begin{figure}
\includegraphics[width=\columnwidth]{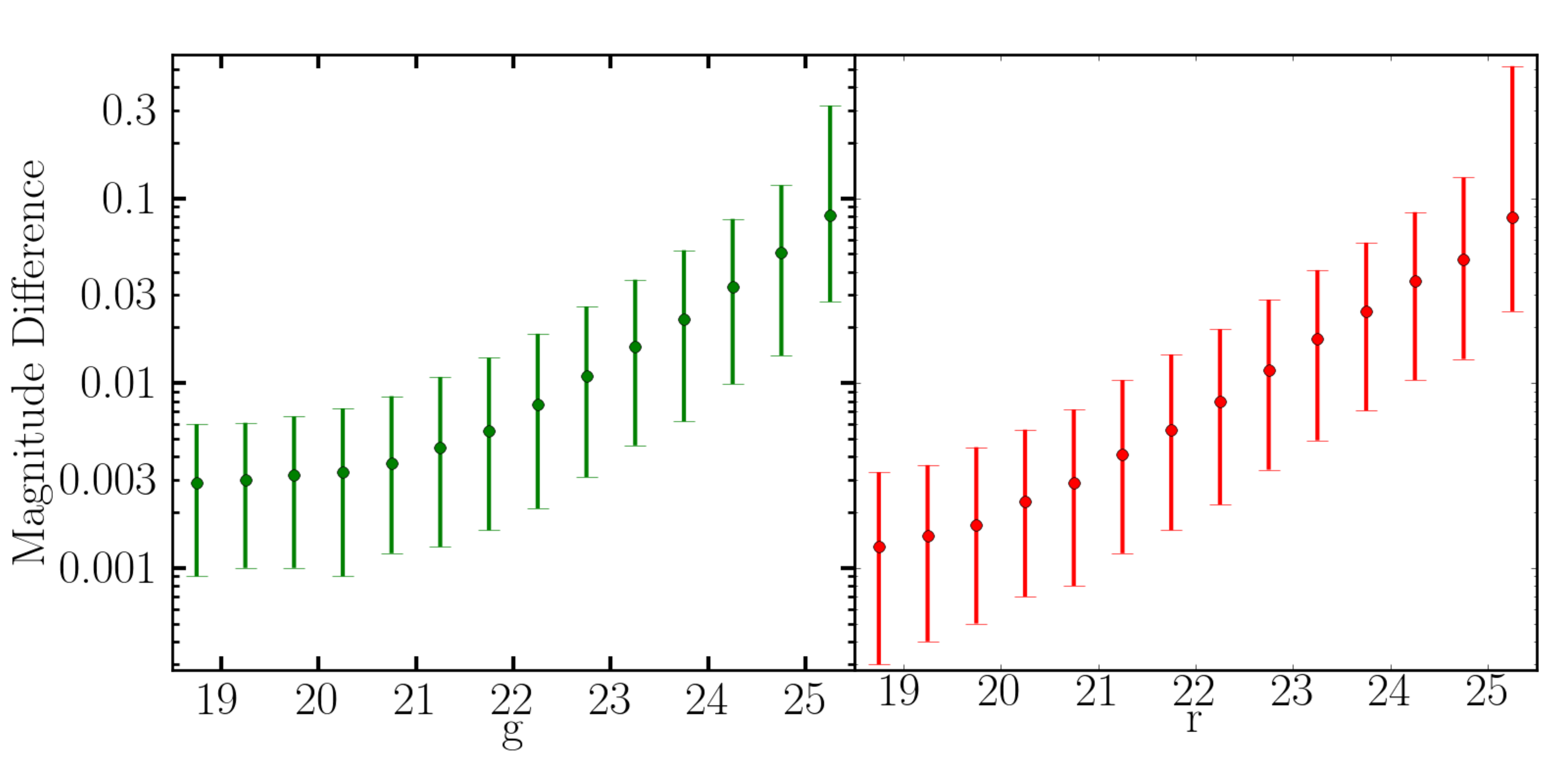}
\caption{ Photometric error v.s. magnitude. Error estimated as the absolute value of the 
difference between input and recovered magnitude for the artificial star test as a function of
input $g-$ (left) and $r-$ (right) magnitudes.
For each magnitude bin we plot the median of the absolute value of the difference along
with the one-sigma interval.}
\label{error_vs_mag}
\end{figure}

To run Talos and calculate the SFH we have to determine a set of parameters associated 
with our data.
These parameters are used to populate the synthetic models that are directly comparable to 
the data.
From the entire CMD we used the photometry of a region that consisted on a magnitude
range of $21$--$24.5$ and a color range of $-0.5$--$1.15$, which we denominated
\emph{photometric CMD region} (blue dashed box on Figure~\ref{cmd_and_starmap}).
The colors and magnitudes of the stars in the photometric CMD region are compared to the 
different models to fit the SFH.
This region was chosen to avoid CMD locations significantly affected by completeness
corrections or contamination sources, while maximizing our signal.
Additionally magnitudes brighter than $g=21$ were excluded to avoid regions of late
stages of stellar evolution, where the different set of isochrones we tried (Teramo,
Dartmouth and Yonsei-Yale) differed significantly from each other.
The individual bins that defined the Hess diagrams were chosen to have a size of $0.1$ in
magnitude and $0.025$ in color.
These dimensions were selected to be small enough to be able to trace the different 
features of the stellar populations in the CMD, and big enough to avoid significant shot 
noise in each bin.
To construct the MDF of the data and the different models, Talos uses a region of the
CMD where significant spectroscopic data exists, which we denominated 
\emph{spectroscopic CMD region} (green solid box on Figure~\ref{cmd_and_starmap}).
In our case this region of the CMD had a $g-r$ color range of $0.5$--$1.25$ and a $g-$ 
magnitude range of $18$--$20.5$.
Then, each time Talos generates a model stellar population it constructs the associated
artificial MDF with the metallicity values of all the stars from the model that lie in the 
spectroscopic CMD region.
Analogously, the MDF of the data is constructed with the stars that have 
spectroscopic metallicity measurements, and are also located within the spectroscopic CMD
region.

The distance to Carina was determined using the magnitude of the RR-Lyrae and red clump 
stars from our data, and we obtained a distance modulus of $20.08\pm0.05$, consistent 
with recent values derived in the literature \citep{weisz14,vandenberg15,coppola13}.
The reddening was calculated on a star-by-star basis, using their equatorial coordinates 
and correlating them to the Schlegel maps \citep{schlegel98a}.
In this way, we correct for differential reddening throughout the galaxy.

To take advantage of the large spatial extent of our data, we determine the SFH
independently for $3$ different concentric regions of Carina:\\
Inner: \,\,\,\,\,\,$0<r/r_{\rm{tidal}}<0.3$\\
Intermediate: \,\,\,\,$0.3<r/r_{\rm{tidal}}<0.6$, and\\
Outer: $0.6<r/r_{\rm{tidal}}<1.3$\\
The number of radial bins was chosen to include enough stars in each region to
enable clear distinction of the different stellar populations in them.
The corresponding dereddened CMDs of the stars from each of these three regions
are shown on the first three panels of Figure~\ref{cmds_diff_regions}, where we can
see systematic differences between them.
For each of these regions, Talos was run using the corresponding Hess diagram (see left 
panels
of Figure~\ref{hess_diagram_fits}) and MDF (blue histograms of Figure~\ref{MDFs}).

\begin{figure}
\centering\includegraphics[width=\columnwidth]{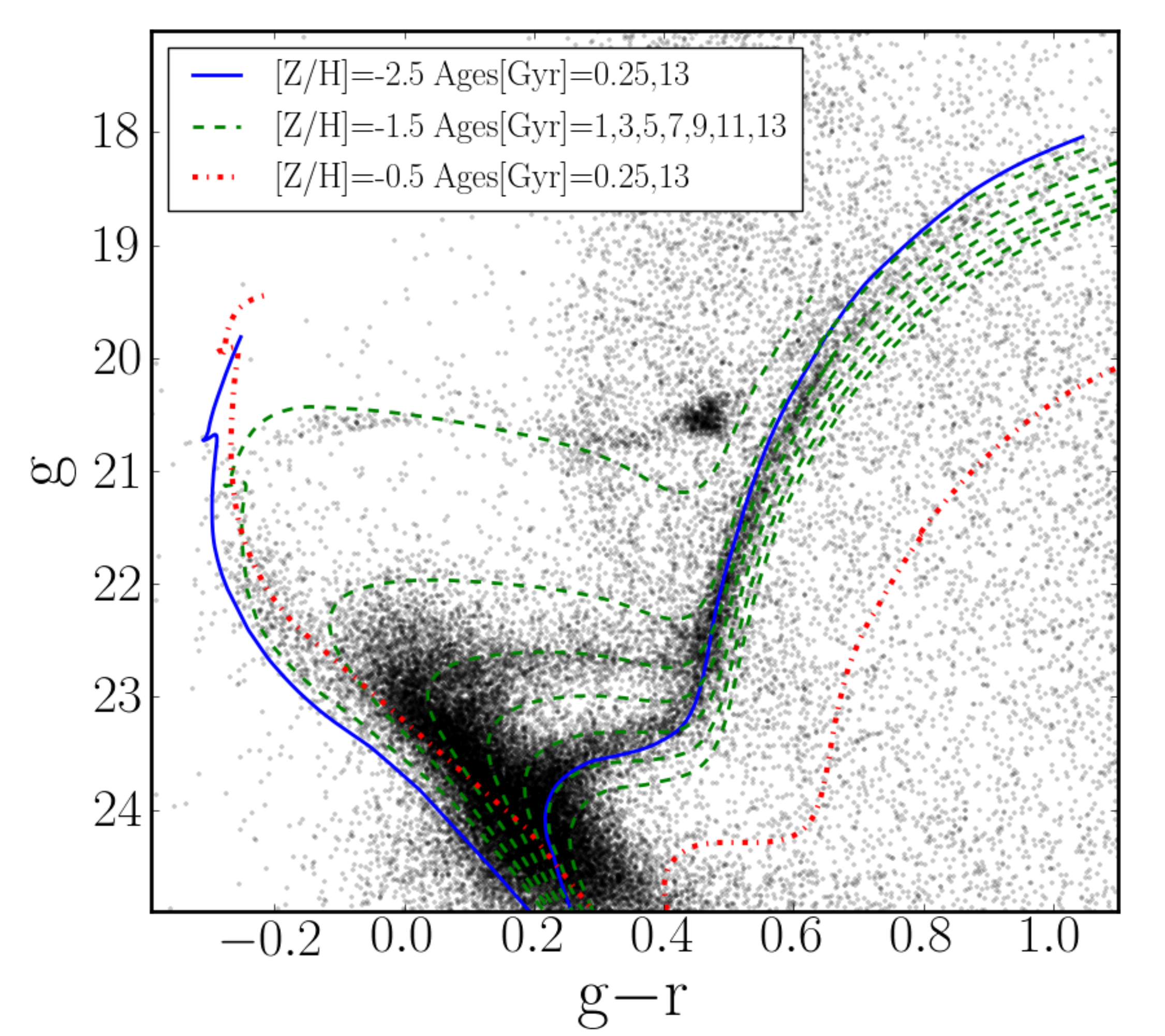}
\caption{CMD of all the stars in Carina used in this study along with a subsample of the
Dartmouth isochrones used to derive the SFH. We show isochrones at three different 
metallicities: $-2.5$,$-1.5$, and $-0.5$. For the isochrones with [Z/H]=$1.5$ we show isochrones
for various ages from $1$ to $13$\,Gyr. For the other two isochrones we just show the lower and 
upper limits of the ages used in this study, which are $0.25$ and $14$\,Gyr respectively.
This plot shows that CMD region of our Carina data is completely contained within the
region spanned by our set of isochrones.}
\label{CMD_with_isochrones}
\end{figure}

To generate the synthetic stellar populations models that are compared to the data, we used 
isochrones from the Dartmouth library \citep{dotter08} which span a wide range of ages,
metallicities and alpha-element abundances. 
The range of metallicities used to define both the SFH and the MDF was 
$-3.5<$[Fe/H]$<-0.5$, which according to Koch's metallicity sample includes all the 
stars in Carina.
The ages used ranged from $0.50$\,Gyr to $14$\,Gyr, which is constrained
by the age of the Universe.
To increase the resolution of the SFH, Talos interpolates between the different isochrones
and after this process we obtained a constant age spacing of $0.50$\,Gyr and a constant 
metallicity spacing of $0.2$\,dex. 
Figure~\ref{CMD_with_isochrones}, shows a subsample of our set of isochrones
over-plotted to the CMD of all the Carina stars we used in this study.
This figure shows that the CMD region populated by the Carina data is completely contained
by the region spanned by our isochrones.
Therefore, 
the different stellar populations conforming the SFH
of Carina have ages and metallicities that are contained within the ranges defined above.

\subsection{Additional input files}

\subsubsection{Contamination}

The Carina dSph galaxy is located at relatively low galactic latitude ($b=-22.2$\,deg) and it
spans a large area in the sky ($r_{\rm{tidal}}=31{\farcm}0$).
Thus, Carina's CMD is significantly contaminated by Milky Way foreground stars.
To analyze the stellar populations in Carina, the foreground stars together with all the 
background sources have to be subtracted from the Hess diagrams.
To account for their contribution, we used the sources located in the region beyond
$2.0\times r_{\rm{tidal}}$ and extending to $3.0\times r_{\rm{tidal}}$, which we 
denominated \emph{contamination region}.
The corresponding CMD of this region is shown on the rightmost panel of
Figure~\ref{cmds_diff_regions}. If we compare this CMD with the ones corresponding
to the different Carina regions, we see that the contribution of Carina's stellar populations
appears negligible in the contamination region.
From the CMD of the contamination region, we constructed a contamination Hess diagram
In this way, the contamination Hess diagram does not include features from Carina's stellar 
populations, and it has enough stars to be a representative sample of the contamination
sources.

To improve the reliability of our contamination Hess diagram, we applied a $6\times 6$ 
Kernel to smooth it.
This significantly reduces the shot noise while still preserving the main features of the 
diagram.
The smoothed contamination Hess was then scaled according to its spatial area and 
subtracted from the original Carina Hess diagrams.
Normalized original and smoothed contamination diagrams are shown in 
Figure~\ref{contamination}, and compared to the Carina Hess diagram.

\subsubsection{Artificial star test}

To calculate the photometric error and the completeness values as a function of 
color/magnitude we ran artificial star tests on our photometry.
We do this by adding stars to our individual images using the IRAF task ADDSTAR.
Then, photometry was carried out following the same procedure used for the original images
to measure the fraction of artificial stars recovered and the magnitude error.
The error was calculated as the difference between the input magnitude and the 
one obtained with DAOPHOT.
For each iteration, only $100$ stars were added to avoid a significant alteration of the 
image crowding.
A total of $280000$ stars were generated, covering all the regions of the CMD and
the spatial extent of Carina inside $1.3\times r_{\rm{tidal}}$.
The distribution of the input magnitudes generated had a maximum at $g\sim25$.
In this way, we increase the resolution of the completeness fraction determinations at the 
magnitudes where these fractions vary sharply.
After performing the photometry on all the artificial stars, we obtained the completeness
fraction as a function of magnitude and color for each region of Carina.
These values were then used to correct the number counts in the Hess diagrams of 
Carina's different regions.
Figure \ref{completeness_vs_mag} shows the completeness fraction as a function of 
$g-$ magnitude (left panel) and $r-$ magnitude (right panel) for three different color 
ranges in the complete Carina region inside $1.3\times r_{\rm{tidal}}$.

The left panel of this figure shows that we reach a $50\%$ completeness at $g\sim24.5$, 
with values at redder colors being slightly higher.
This implies that we have completeness fractions higher than $50\%$ for the entire
photometric CMD region defined in \S3.2.
With the artificial star test, we also estimated the magnitude error of our 
photometry, calculated using the absolute value of the difference between input and 
recovered magnitudes of the artificial stars. The median of this value was calculated for
each magnitude bin, and these numbers were then used to reproduce the photometric
errors on the synthetic stellar populations.
Figure~\ref{error_vs_mag} shows the median and one-sigma intervals of the magnitude error
estimation as a function of $g-$ (left) and $r-$ (right) magnitudes.

\section{Results}

\begin{figure*}
\centering%
\includegraphics[height=0.8 \textheight]{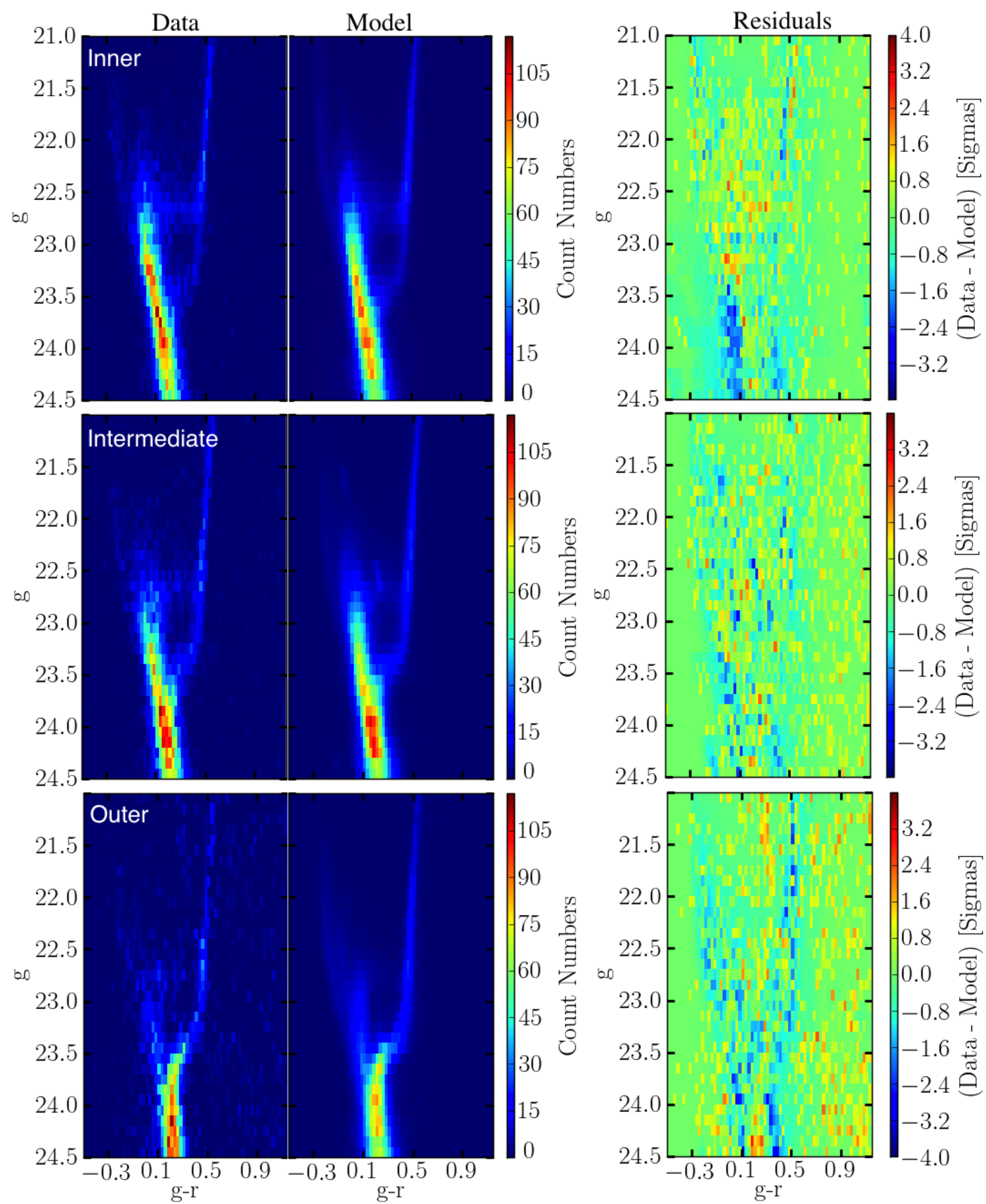}
\caption{\small Comparing observed and model CMDs represented by Hess Diagrams. Colors
in plots represent number of stars in each CMD region. Each panel shows at left the CMD of
the data, in the middle the best fit model CMD, and at the right the residuals calculated as 
(model$-$observed)/errors.
Top: inner region corresponding to $0<r/r_{\rm{tidal}}<0.3$.
Middle: intermediate region corresponding to $0.3<r/r_{\rm{tidal}}<0.6$.
Bottom: outer region corresponding to $0.6<r/r_{\rm{tidal}}<1.3$.
}
\label{hess_diagram_fits}
\end{figure*}

The SFH for each of the Carina regions defined in \S3.2 was determined 
by comparing our data with a set of synthetic stellar populations, which were created
using a set of Dartmouth isochrones.
To generate the models that are compared to the data, Talos 
uses the binary fraction of the
system as an input.
Since this number has not been determined observationally for Carina, we allow
the code to find the best choice. For this,
we performed various runs of Talos, using different binary fractions each time.
After this, we chose a binary fraction value\footnote{Even though the
binary fraction is, in principle, a function of the stellar mass \citep[e.g.,][]{kouwenhoven09},
the vast majority of the stars we are using to fit the SFH in Carina are close to the
main-sequence turnoff and therefore, their masses are closely clumped in the range
$0.7$--$0.9$\,M$_{\odot}$.} 
for Carina of $0.4$, because this value produced the Hess 
diagram models that most resembled the data.
This binary fraction is consistent with values obtained in other dwarf galaxies
\citep[e.g.,][]{geha13a,mcconnachie10}.

Figure~\ref{hess_diagram_fits} shows the Hess diagram of the data (left panels) and the best fit 
model (right panels)
as well as the residuals for each Carina region.
We see in this figure that for all regions, the best fit is an accurate representation of the 
observations, with residuals lower than one sigma for $80$\% of the color/magnitude bins 
and lower than $2$\,sigma for $\sim99$\% of the bins.
The largest systematic differences from the data and best models lie in the main sequence
color for the inner region, which is bluer for the model than for the data, and the
main-sequence width for the outer region, which is larger for the model than for the data. 
The reason for these systematics may be attributed to a combination of an offset between the assumed 
binary fraction/distribution and the real one, an incorrect estimation of the magnitude error as a function 
of magnitude, uncertainties in the calibration of the data, and an error in the isochrone models
used.
Even though the specific origin of our systematics could not be identified, 
we do not deem them significant
since they are in almost all cases similar to the expected
random errors.

The model Hess diagrams reveal 
the presence of an old and an intermediate-age population for all Carina regions.
But the relative importance of the intermediate-age population decreases as we move to
larger radii.
A combination of these two populations correctly reproduce the main features of Carina's CMDs,
such as main-sequence turnoff positions, the gap separating the two main sub-giant 
branches, and the width of the RGB.
There are also hints of a young population in all of Carina's regions, accounting for
$1$ to $2$ percent of the stellar mass, and their authenticity
will be analyzed in \S5.

\begin{figure}
\centering%
 \includegraphics[width=\columnwidth]
{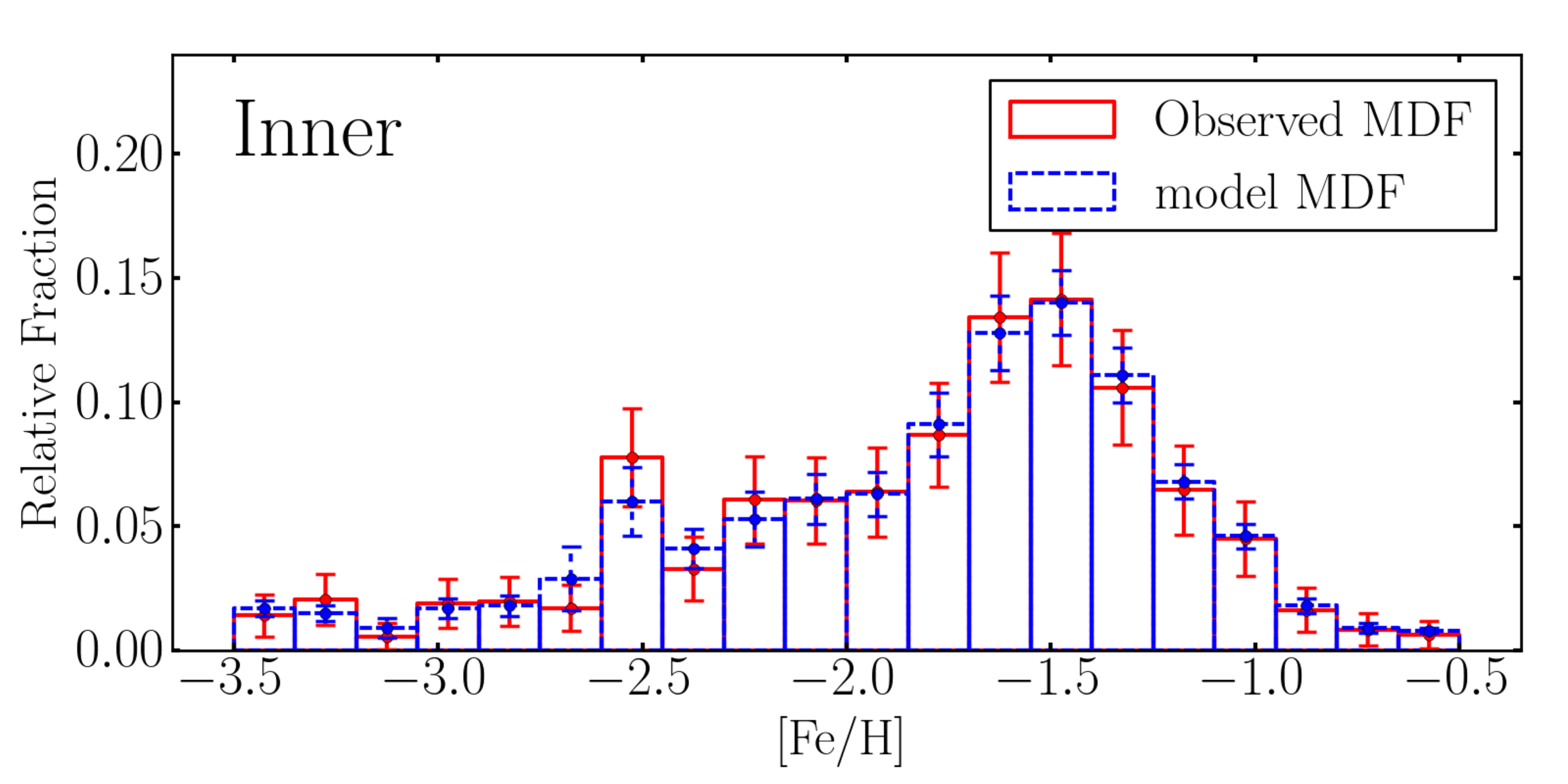}\\
\includegraphics[width=\columnwidth]
{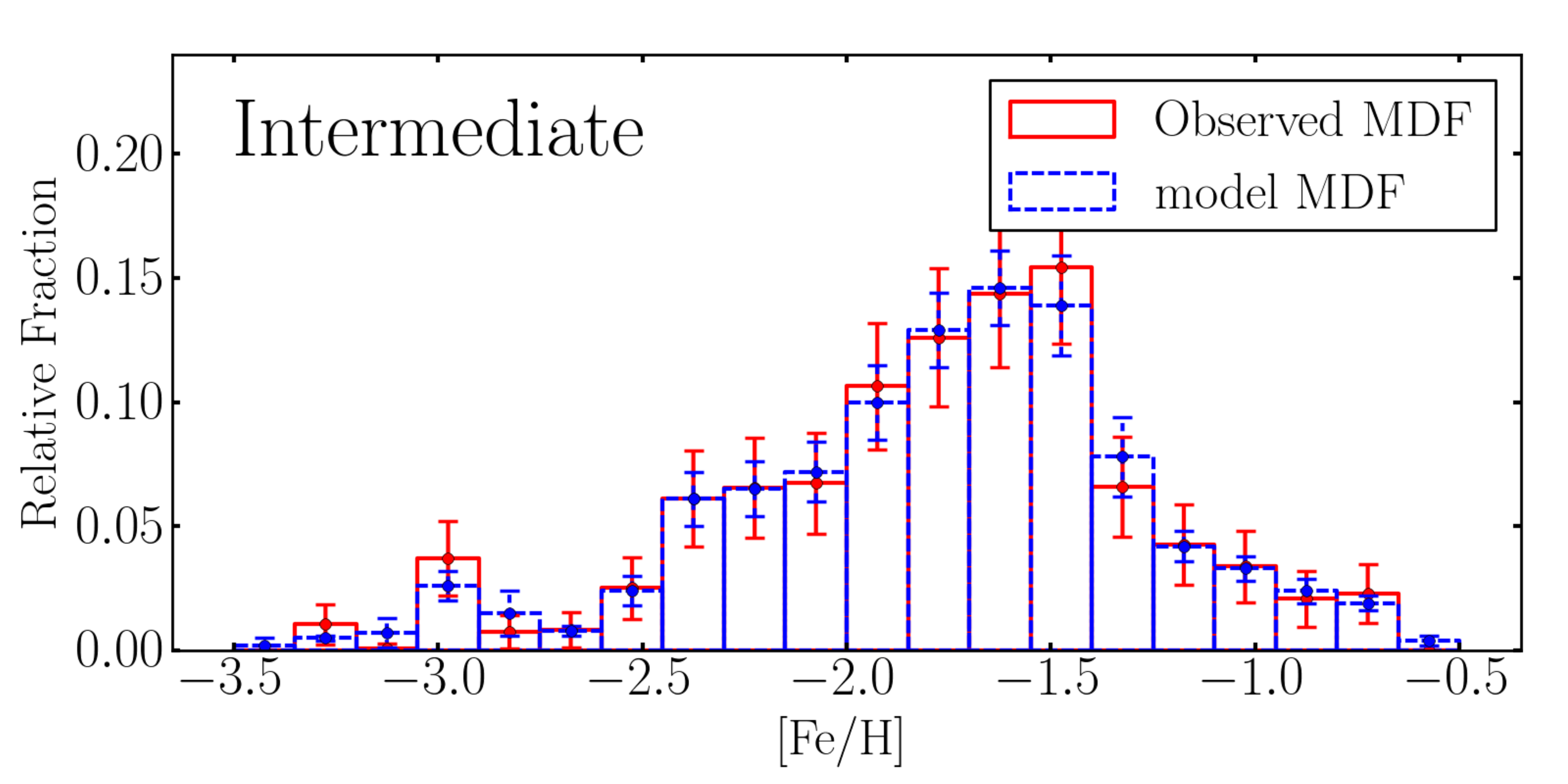}\\
\includegraphics[width=\columnwidth]
{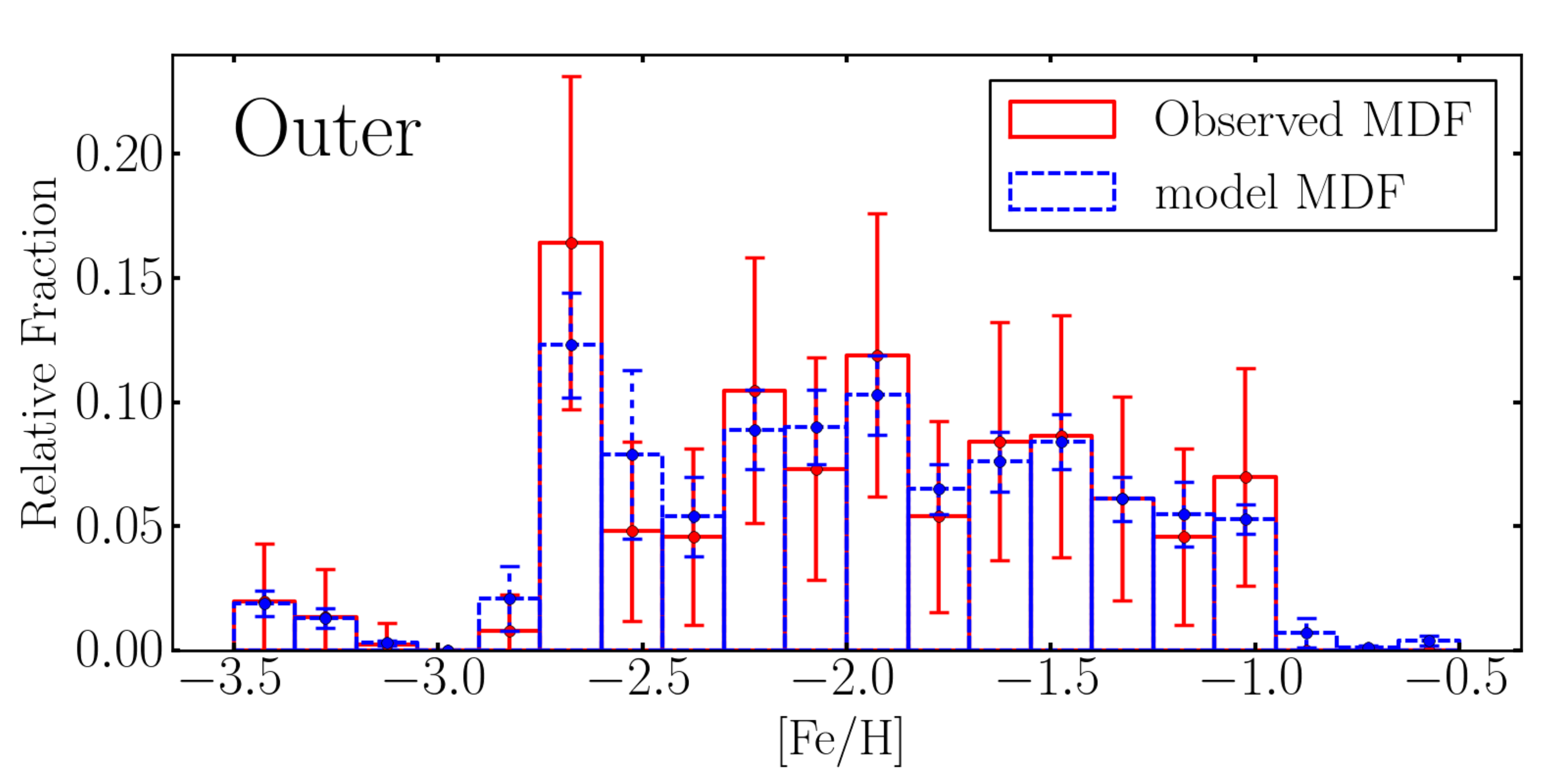}\
\caption{ MDF of observations and model in different regions
of Carina. 
These distributions include only the stars from spectroscopic CMD region (defined in
Section~3.2).
Top: inner region corresponding to $0<r/r_{\rm{tidal}}<0.3$.
Middle: intermediate region corresponding to $0.3<r/r_{\rm{tidal}}<0.6$.
Bottom: outer region corresponding to $0.6<r/r_{\rm{tidal}}<1.3$.
}
\label{MDFs}
\end{figure}

\begin{figure}
\centering%
\includegraphics[width=\columnwidth]{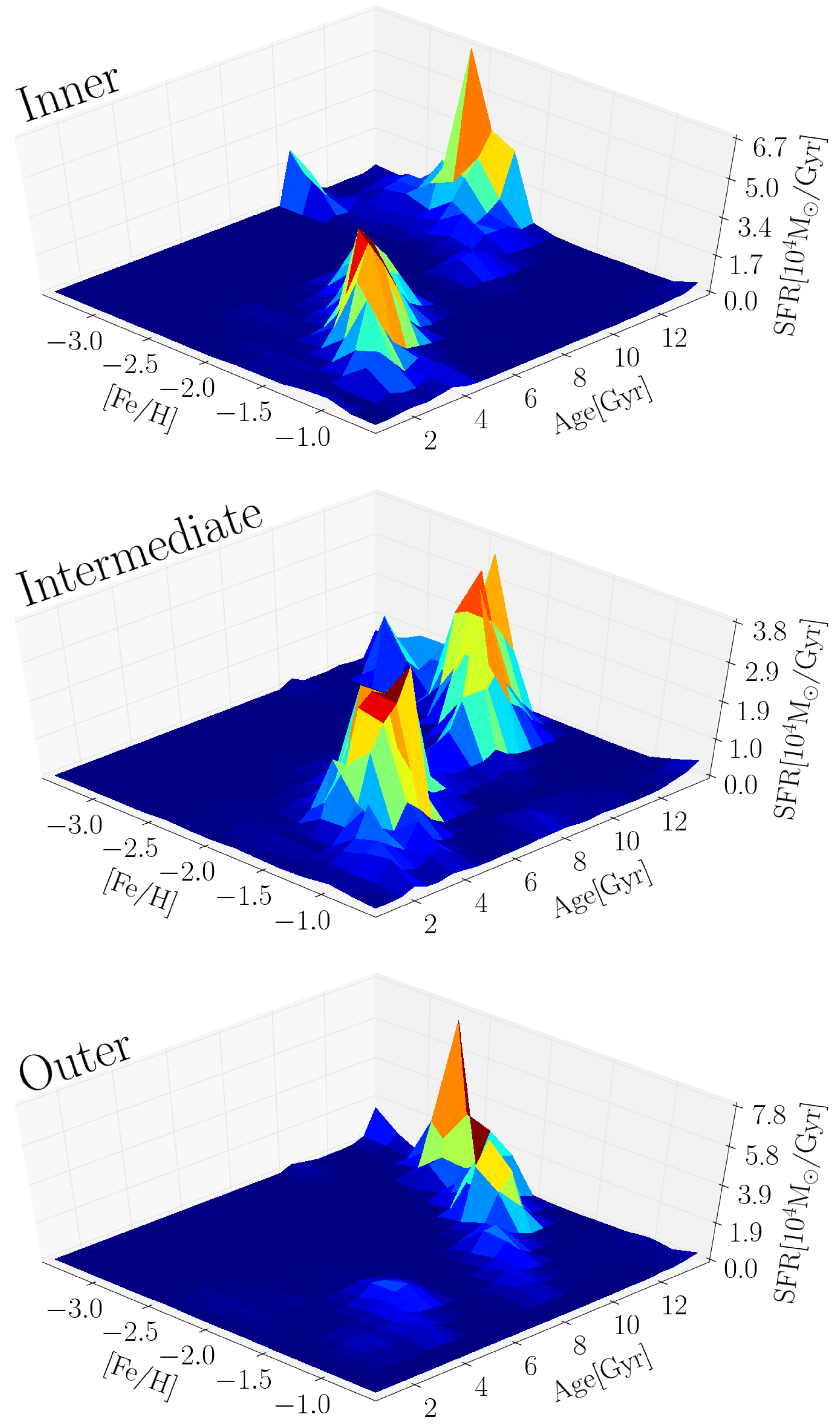}%
\caption{ 3D SFH of different regions of Carina.  Color and height of each
  bin shows the stellar mass formed at that age and metallicity.  Top:
  inner region corresponding to $0<r/r_{\rm{tidal}}<0.3$.  Middle:
  intermediate region corresponding to $0.3<r/r_{\rm{tidal}}<0.6$.  Bottom:
  outer region corresponding to $0.6<r/r_{\rm{tidal}}<1.3$.  }
\label{3D_SFHs}
\end{figure}

\begin{figure}[htp]
\centering%
 \includegraphics[width=9.0cm]{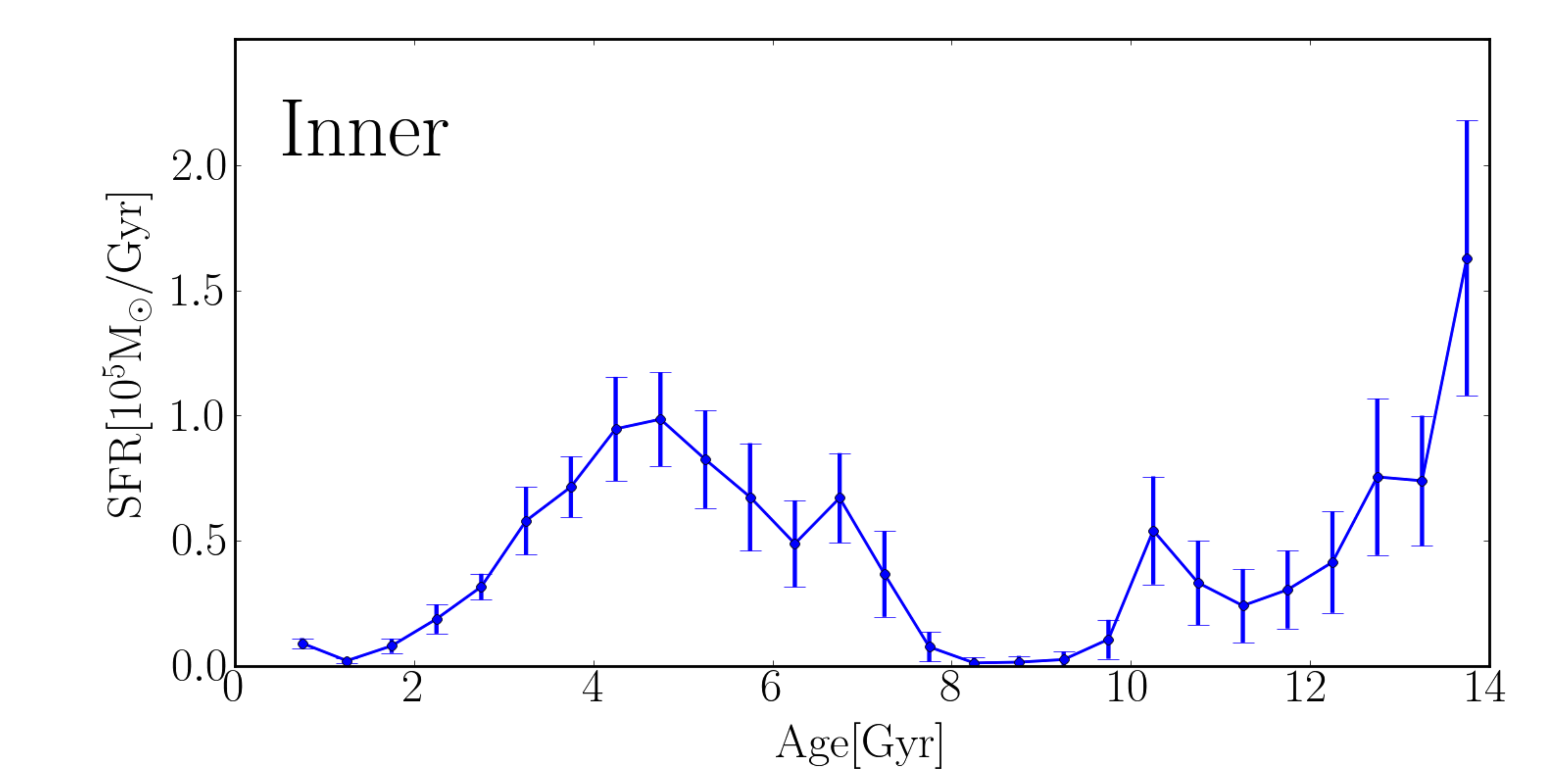}\\
\includegraphics[width=9.0cm]{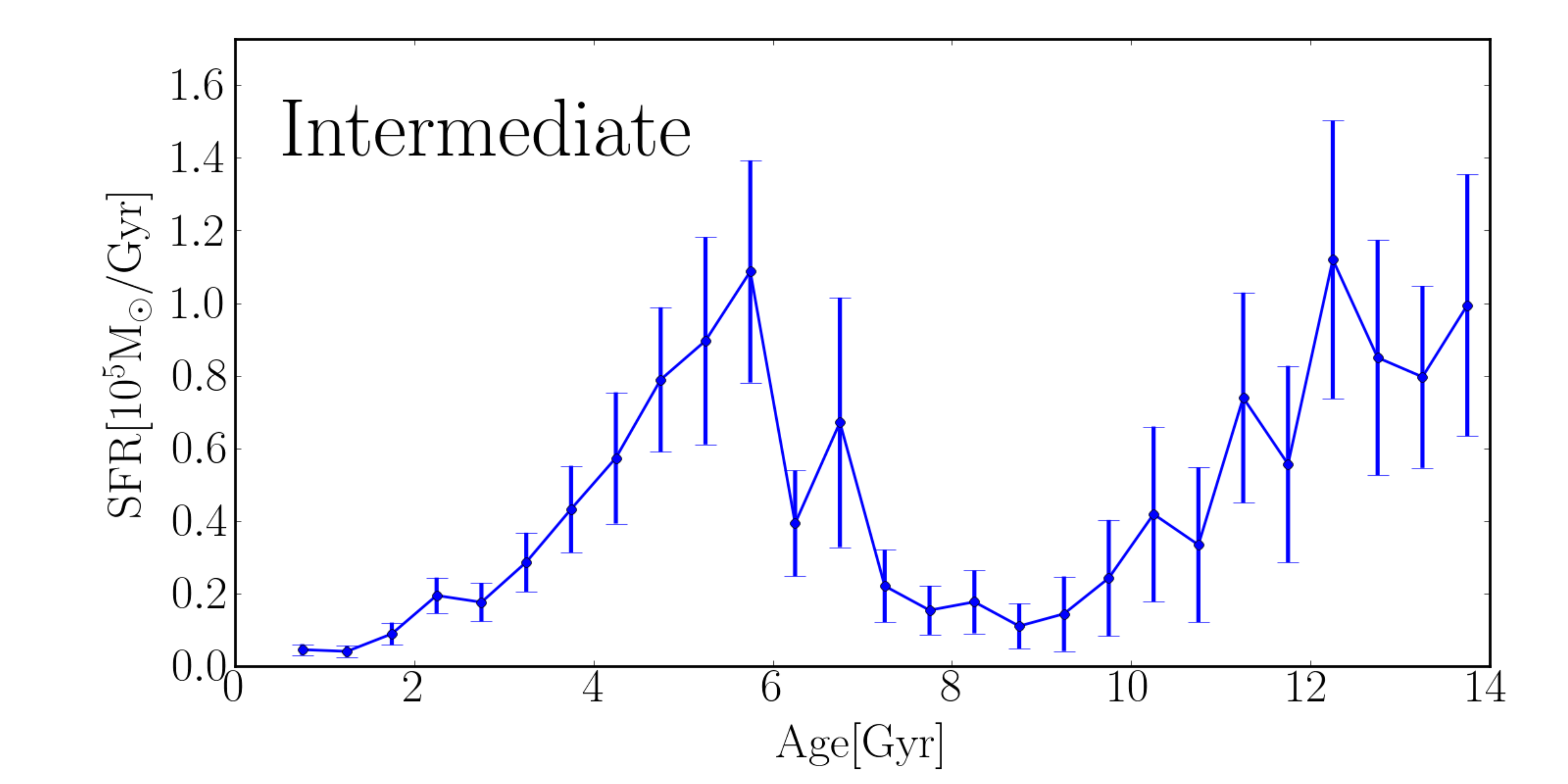}\\
\includegraphics[width=9.0cm]{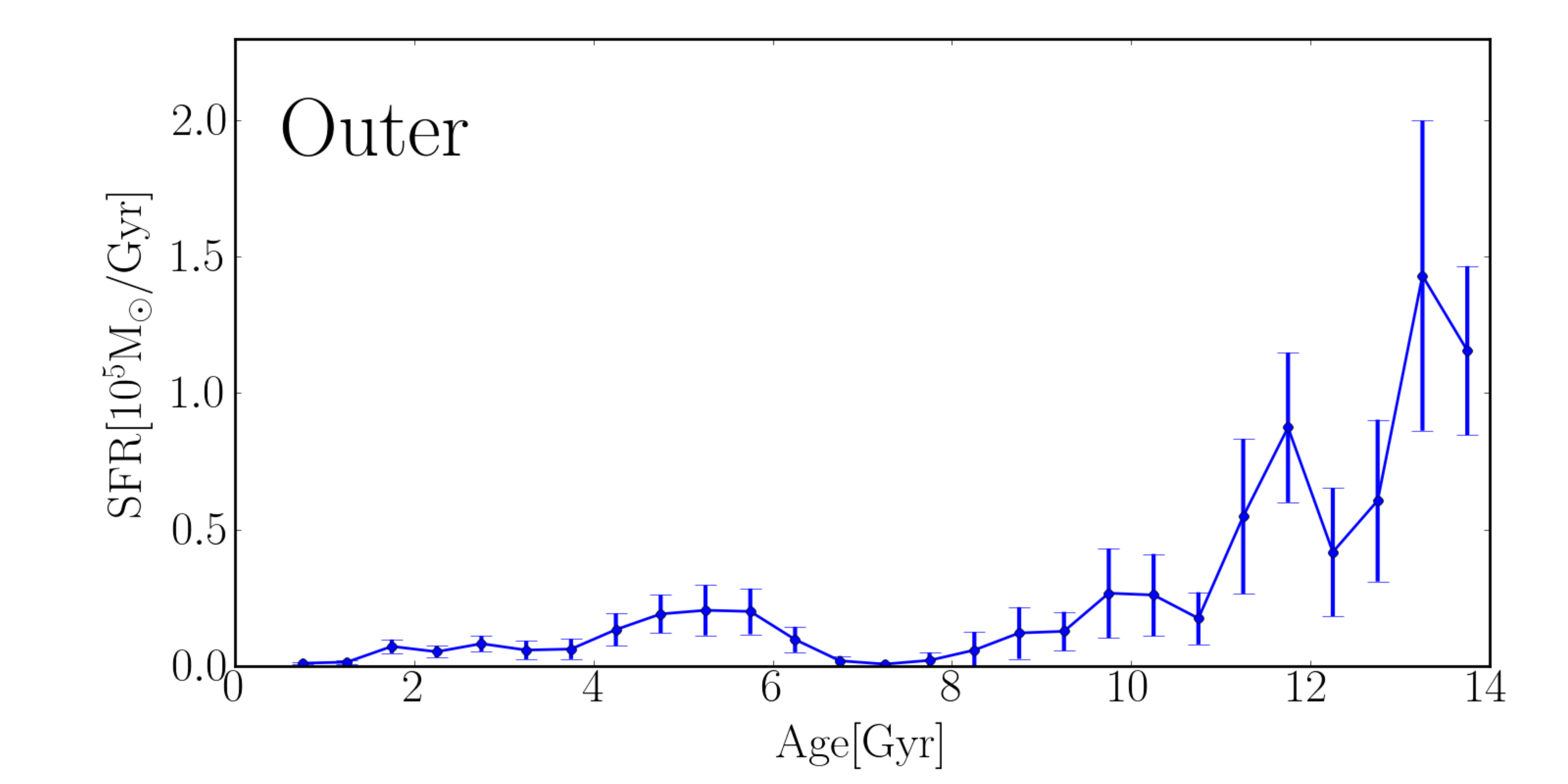}\
\caption{ Star formation rate as a function of age for different regions of Carina.
Top: inner region corresponding to $0<r/r_{\rm{tidal}}<0.3$.
Middle: intermediate region corresponding to $0.3<r/r_{\rm{tidal}}<0.6$.
Bottom: outer region corresponding to $0.6<r/r_{\rm{tidal}}<1.3$.
}
\label{SFHage}
\end{figure}

\begin{figure}[htp]
\centering%
 \includegraphics[width=9.0cm]{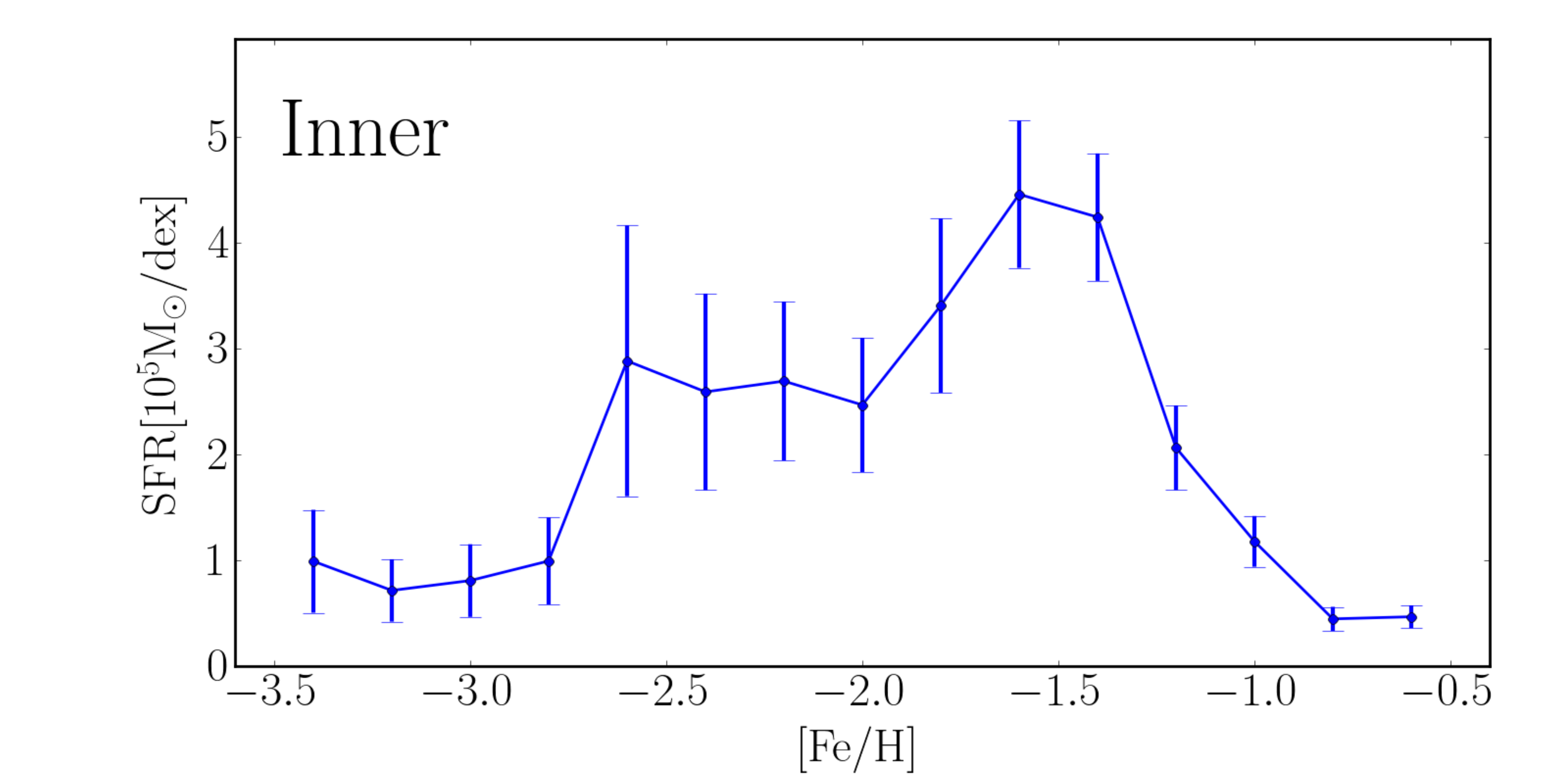}\\
\includegraphics[width=9.0cm]{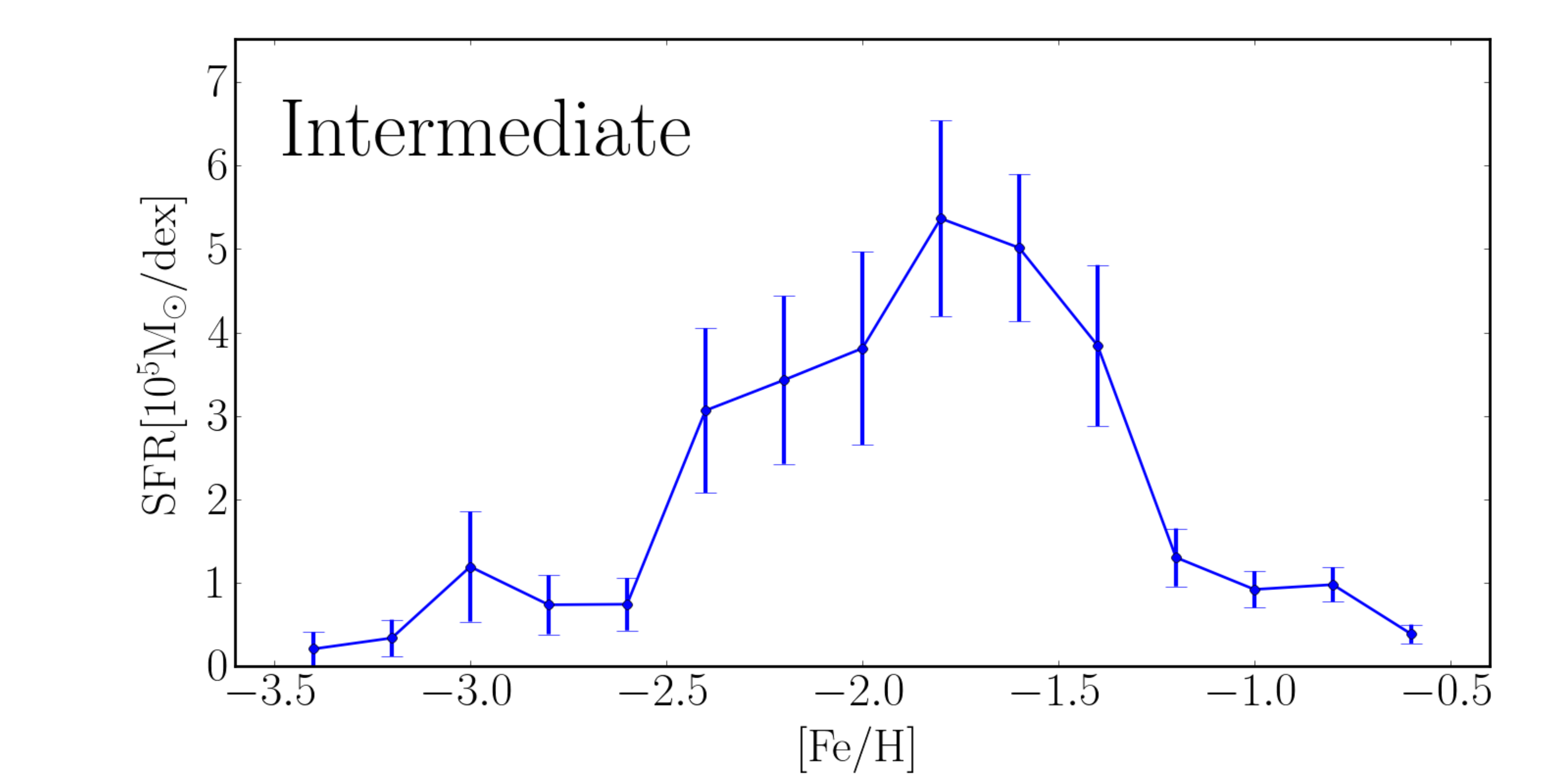}\\
\includegraphics[width=9.0cm]{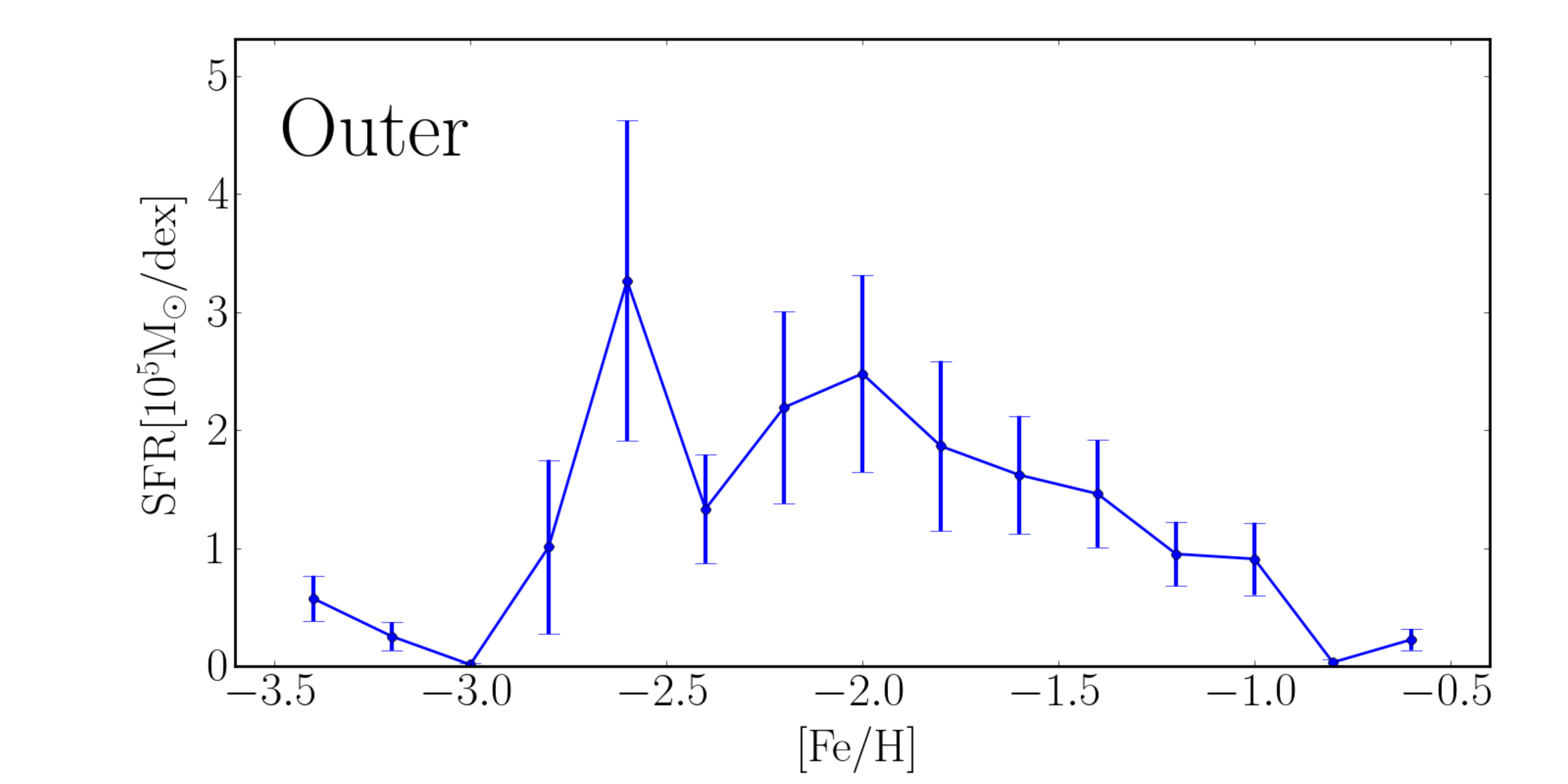}\
\caption{ Star formation rate as a function of metallicity for different regions of Carina.
Top: inner region corresponding to $0<r/r_{\rm{tidal}}<0.3$.
Middle: intermediate region corresponding to $0.3<r/r_{\rm{tidal}}<0.6$.
Bottom: outer region corresponding to $0.6<r/r_{\rm{tidal}}<1.3$.
}
\label{SFHmet}
\end{figure}

Figure~\ref{MDFs} shows the comparison between data and best-fit model for the MDFs.
The histograms in this figure include only stars from the spectroscopic CMD region
(defined in \S3.2).
It is important to note that the model histograms in this figure (blue histograms) are 
different from those in Figure~\ref{SFHmet}, because the MDFs presented in those plots 
include the stars in the model from the entire CMD region.

MDFs of the models closely reproduce the observations for all Carina regions, with
differences between the two always within the uncertainties.
This figure indicates that the average metallicity of the RGB stars in Carina decreases with 
radius, showing a large spread at all regions.

The full SFH derived for each Carina region is shown in Figure~\ref{3D_SFHs}.
This plot displays the stellar mass formed at each age/metallicity combination and clearly
shows the presence of two main episodes of star formation separated by a star formation 
temporal gap.
The old episode occurred more than $10\,$Gyr ago and its stars have very low
metallicity ($-3.0<$[Fe/H]$<-2.0$).
The intermediate-age episode started $\sim8\,$Gyr ago and stopped $\sim2\,$Gyr ago,
increasing its metallicity from $\sim-2.0$ to $\sim-1.0$ in that period.
This shows that the average stellar metallicity at the moment the intermediate-age episode 
began is consistent with the stellar metallicity of the last stars formed in the old episode,
and therefore there is a continuous transition in stellar metallicity between the two
episodes.
The temporal gap of star formation separating the two main bursts represents an epoch 
where no significant star formation took place.
This gap lasted for a few billion years between $\sim10$ and $8\,$Gyr ago.

The central age and metallicity of the two main episodes is consistent through all Carina
regions, however, their relative importance varies from inner to outer regions.
The inner region is dominated by the intermediate-age episode while still having an 
important contribution from the old one.
The middle region displays both episodes with similar relative importance.
Finally, the outer region is composed mainly by old stars with a very small contribution 
from the intermediate-age episode.

There is no clear evidence for a young (age $< 2\,$Gyr) episode in Figure~\ref{3D_SFHs}, 
which could indicate that there are no young stars in Carina, or that their 
contribution is negligible compared to the two main episodes.
The presence of a young population has been claimed by previous studies
\citep[e.g.,][]{monelli03,hurley-keller98}, and hence, in the following section we will
analyze the authenticity of the young population in Carina and the distinction between
these stars and blue stragglers.

As shown by \citet{aparicio09}, the main source of statistical uncertainties on  
SFH results derived from the synthetic CMD method, is the specific choice of
the grids that define the Hess diagrams, MDFs, and the resulting SFHs.
The general hypothesis is that there is a loss of information produced by discretizing the
values contained in these distributions, according to the arbitrary grids defined by the user.
To account for this source of error, for each of the regions we defined for Carina (inner, 
intermediate, and outer), we carried out $32$ runs of Talos in it.
Each time a trial of Talos was performed, we shifted the grid defining the input 
photometry/spectroscopy grid and/or the grid defining the resulting SFH.
Then the final SFH of a given Carina region was determined as the average of the solutions
obtained in all the trials, while the associated errors were calculated as the standard
deviations of the different solutions.

The SFH as a function of age is shown in Figure~\ref{SFHage}, where we can see the same 
general features as in Figure~\ref{3D_SFHs}.
There are two main episodes separated by a temporal gap in star formation and the relative 
importance of the old episode increases with radius, indicating a positive age radial 
gradient present in Carina, which is steeper between the middle and outer regions.
Figure~\ref{SFHmet} shows the SFH as a function of metallicity (or chemical enrichment 
history).
In these plots we can see that the metallicity distribution is similar in the inner and
intermediate region, but the average metallicity in the outer region is systematically
lower than in the rest of the galaxy, indicating a negative metallicity gradient.
From the results presented in this figure, we calculated that Carina has a mean metallicity
of [Fe/H]=$-1.81\pm0.29$ and a spread of $\sigma =0.54$\,dex.
The SFH of Carina that we derived is able to reproduce both the
large dispersion MDF and 
the narrow RGB seen in the Hess diagrams.
This shows that these two features in Carina are consistent and their implications will be
interpreted in the following section.
From the SFH derived for Carina we also estimated the total stellar mass formed in the
galaxy, obtaining $1.60\pm0.09\times 10^{6} \, M_{\rm{\odot}}$ within 
$1.3\times r_{\rm{tidal}}$, and $1.45\pm0.12\times 10^{6} \, M_{\rm{\odot}}$ within
the nominal $r_{\rm{tidal}}$. The latter value is larger than previous values derived for Carina
(e.g., \citealt{deBoer14} derived a value of $1.07\pm0.08\times 10^{6} \, M_{\rm{\odot}}$ 
within the nominal $r_{\rm{tidal}}$).
With these results we also estimated the total visual luminosity of Carina as 
L$_{\rm V}=7.3\pm0.4\times 10^5\,$L$_{\rm \odot,V}$.

\section{Discussion}

\begin{figure}
\includegraphics[width=\columnwidth]{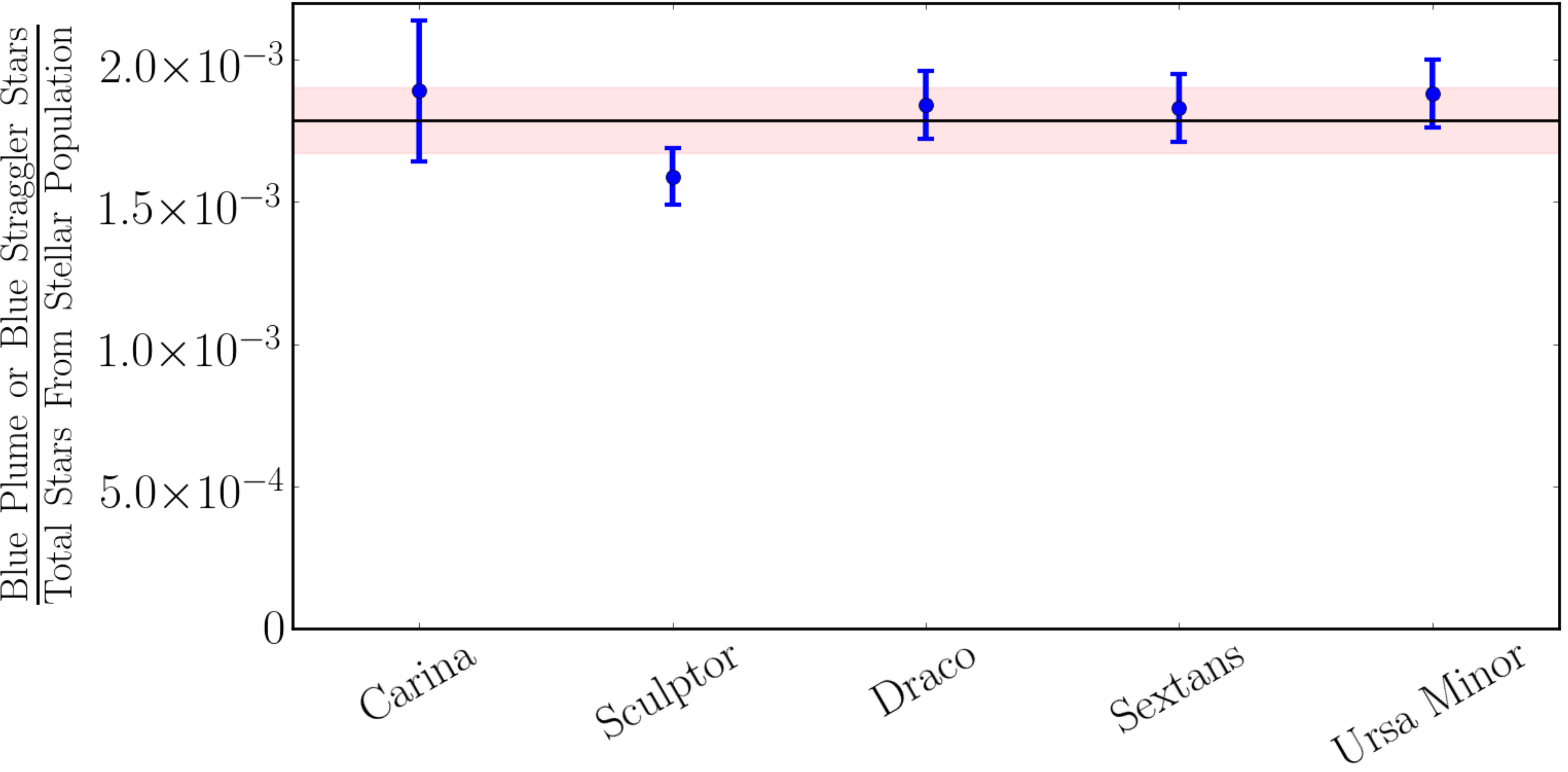}
\caption{Blue straggler/blue plume fractions for the different galaxies.
For Carina, we show the number of blue plume stars over the number of RGB
stars from the intermediate-age population. These blue plume counts include, in principle, 
young stars and blue stragglers formed from the intermediate-age stellar population.
For all the other galaxies we show the blue straggler fraction calculated as the number of
blue stragglers over the number of RGB stars. Black solid line indicates the average
blue straggler fraction for all the galaxies, and the red region shows the area located at a 
maximum of 1 standard deviation difference from the average.}
\label{BSS_trend}
\end{figure}

\subsection{Young Population v.s. Blue Straggler Population}

The star formation of Carina clearly shows at least two different episodes.
Even though a young population is not clearly evident, as seen in  
Figures~\ref{hess_diagram_fits}, \ref{3D_SFHs} and \ref{SFHage},
a small but not negligible contribution of young stars (age$<2$\,Gyr) might also be present 
in Carina, which can account for $1$--$2\%$ of the total stellar mass according to our SFH
derivation.
Different studies have claimed the presence of a young population in Carina
\citep[e.g.,][]{hurley-keller98,mateo98b,monelli03}. However,
there exists the possibility that these stars correspond to blue
stragglers mistakenly classified as young stars. 
Blue straggler stars are coeval to a given stellar population, but bluer
and brighter than its main sequence turnoff.
Thus, these stars can populate the same region as young stars in a CMD, and therefore,
their differentiation is not straightforward.

To analyze the authenticity of the young population of Carina, we developed a method
to discriminate blue straggler counts from young star counts.
To apply this method, we start by calculating the number of stars in Carina located in a 
region brighter and bluer than the main sequence turnoff of the intermediate-age 
population.
In principle, we do not know if these stars are blue stragglers or young stars, and thus, we 
will denominate the sum of both contributions as blue plume counts.
The goal is to compare the blue plume fraction of Carina with the blue straggler
fraction found in a subsample of the local dSph galaxies used in \citet{santana13a}.
For those galaxies, the authors showed that the stars brighter and bluer than the
main-sequence turnoff are genuine blue stragglers. Additionally it was shown that local
dSphs share a common fraction of blue stragglers stars \citep{momany07a,santana13a}.
Therefore, by comparing the blue plume fraction of Carina with the blue straggler fraction
in these local dwarf galaxies, we aim to check what contribution of the blue plume
fraction of Carina comes from blue stragglers and what fraction from young stars.
Using this result, if the fraction of stars in the blue plume of Carina is similar to the blue
straggler fractions in the other local dSphs, then they are most likely dominated by blue
stragglers.
But if this fraction is considerably higher than the average blue straggler fractions in other
dSphs, then the blue plume is probably dominated by young stars. 
Details on how the blue plume and blue straggler fractions were calculated are given in the
Appendix.

The final blue straggler fraction for each of the comparison local dSphs is shown in
Figure~\ref{BSS_trend}, along with the blue plume fraction found for Carina.
As we can see in this plot, local dSphs share a common blue 
straggler fraction \citep[as found in ][]{santana13a}.
It can also be seen in Figure~\ref{BSS_trend}, that the difference between the blue plume 
fraction of Carina and the blue straggler fraction in our dwarf galaxies, is less than the
standard deviation of the blue straggler fraction.
This shows that the number of stars in the blue plume of Carina's CMD is
completely consistent with the number of blue stragglers found in other local dSphs;
and hence, we claim that there is no considerable contribution of young
($<2\,$Gyr) stars in Carina.
We thus conclude that the Carina dSph galaxy had only two episodes of star formation
and the small contribution of young stars derived by Talos are due to misclassified blue 
straggler stars.

It is worth noting that genuine young stars have been claimed in previous studies about
the SFH of Carina \citep[e.g.,][]{monelli03,hurley-keller98}.
However, the relative fraction that these stars represent from the total stellar mass of
the system was not quantified in these studies.
Furthermore, the arguments used by these authors to claim that these stars are young
and not blue stragglers are based on the hypothesis that these stars are much more
numerous than the expected number of blue stragglers in \emph{globular clusters}.
There are two possible problems in associating the properties of blue stragglers in globular
clusters to the blue stragglers in the Carina dSph galaxy.
First, \citet{santana13a} showed observationally that the fraction of blue stragglers in dwarf 
galaxies can be of the order of $10\times$ higher than those in globular clusters, which
according to the authors, is because of the higher prevalence of binaries from which blue
stragglers are formed in the low density environments present in local dSph galaxies.
Second, Milky Way globular clusters are old stellar populations (older than 
$\sim10$--$11$\,Gyr according to recent reviews such as \citealt{gratton12}) whereas
Carina contains an old and an intermediate-age episode.
Main-sequence stars from the intermediate-age population have a range in 
luminosity and mass that extends to higher values than the one of old stars.
Therefore, the merger or mass transfer between the intermediate-age stars from Carina
should produce blue stragglers brighter than the ones formed in globular clusters.
Thus, we claim that the number of blue stragglers in Carina brighter
and bluer than the intermediate-age main-sequence turnoff, is not expected to match
the properties of blue stragglers in globular clusters.

Further complementary observations can significantly help to confirm the nature of the
stars in the blue plume of galaxies like Carina.
For example, \citet{gosnell14} have confirmed the presence of $3$ hot and young White 
Dwarfs as companion stars in binary systems for three blue straggler stars in the open
cluster NGC~188.
These White Dwarfs would correspond to former RGB or AGB stars that transferred mass in
the recent past to their companion stars in the binary system, and hence, transforming
them into blue stragglers.
These stars were detected as a high-significance UV excess
using three narrow band filters from the Advance Camera Survey on board HST.
The spectral energy distribution derived from the photometry of these stars, was well fit by
the sum of the contributions of a single blue straggler and a While Dwarf of different
temperatures for the three cases.
That study represents the first direct determination of the formation mechanism for
individual blue stragglers, and further observations of this type can significantly help to
confirm or rule out the blue straggler nature of different blue plume populations.

\subsection{SFH of Carina: Internal Evolution v.s. External influence}

Carina is a local dSph galaxy whose SFH has raised great interest since its 
discovery \citep{cannon77}, mainly because it is the only dSph in the Local Group 
showing clear multiple main-sequence turnoffs.
Additionally, Carina displays signs of tidal influence from the Milky Way 
\citep{munoz06b,battaglia12}. In this context, one of the key questions regarding Carina's 
SFH is how much it is governed by interactions with the Milky Way as opposed to being the
result of internal evolution.

The depth and quality of the data presented in this work represent a key opportunity for 
unraveling the nature of Carina's SFH. We have $g-$ and $r-$ Sloan band photometry 
reaching a $50\%$ completeness at g$\sim24.5$, more than one magnitude fainter 
than the oldest turnoff.
In addition, the field of view of our observations encompasses an area of 
$\sim 2$\,deg$^{\rm{2}}$, which translates into full coverage out to
$1.3\times r_{\rm{tidal}}$, and partial coverage to $3.0\times r_{\rm{tidal}}$.
The combination of depth and coverage of the data is important for tracing the 
oldest stellar population which extends even farther than $r_{\rm{tidal}}$ and has a faint 
main-sequence turnoff.

In this work, we have used the full Hess diagrams of different regions in Carina along with the 
MDF from public spectroscopy \citep{koch06}, and compared them to synthetic stellar 
populations using the routine Talos \citep{deBoer12}.
With this method, we obtained a SFH composed of two episodes separated by a temporal 
gap in star formation.
The episodic behavior of the SFH of Carina is consistent with what had been previously
claimed by various previous studies
\citep[e.g.,][]{smecker-hane96,hurley-keller98,deBoer14}.
The first episode corresponds to an old (age$\sim10$--$13.5$\,Gyr) population with 
metallicities in the range of [Fe/H]$\sim-3.0$--$-2.0$, and the second one is an
intermediate-age ($2$--$8$\,Gyr) episode with metallicity increasing with age from
[Fe/H]$\sim -2.0$ to [Fe/H]$\sim -1.0$.
The old and intermediate-age episodes correspond to $54\pm4\%$ and $45\pm4\%$
of the stellar mass respectively.
This is the first study of the SFH of Carina where the old episode is found to be 
the majority of the population.
This result is in agreement with what is often found for other local dSphs
\citep[e.g.,][]{grebel99a,tolstoy09}.
The different relative fractions obtained for the old population of Carina might be a result of
our deeper data.
The proper photometric characterization of the old main-sequence
turnoff is given by the fact that it is more than one magnitude brighter than the $50\%$
completeness level of our photometry (see Figure~\ref{cmd_and_starmap}).
This has a stronger effect in the 
outermost regions of Carina where the old
population is relatively more important, i.e., 
we obtained higher relative fractions for the old population than previous studies
simply because we can detect them more reliably and to larger radii.
Considering that Carina might have experienced signifiant mass loss in the past due to tidal
influence \citep[e.g.,][]{munoz06b} and that this influence is stronger for more extender
stellar components \citep{sales10}, then the older stars would have preferentially been
removed from Carina. Given this, the fraction of old stars presented in here would represent
the \emph{current} value, and a lower bound to the fraction of old stars formed in the
galaxy. If this is the case, then the difference between the number of stars from the old
and intermediate-age populations would be even larger, and the prevalence of the old
stellar population would be clearer.
Separating these two main episodes there is an epoch consistent with no star formation at 
all, occurring between $8$ and $10$\,Gyr ago, 
previously observed by various
studies \citep[e.g.,][]{hurley-keller98,bono10,deBoer14}.
Finally, 
the data show evidence that could be interpreted as a small contribution of young stars 
(less than $2\%$ of stellar mass).
However, as we concluded in Section~5.1, we believe that these stars are most likely 
dominated by blue stragglers.

The SFH derived for Carina in this work produces good matches with both the input CMDs 
and MDFs as shown by Figures \ref{hess_diagram_fits} and \ref{MDFs} respectively.
This implies for example, that a metallicity distribution with a large dispersion and a narrow
color distribution can be completely consistent with a complex SFH, with two
important episodes separated by a couple of Gigayears which indicates a large
metallicity-age degeneracy.

If we analyze the SFH of the inner, intermediate and outer region of Carina separately, we
see that all the regions show the same central ages and metallicities for the two episodes.
However, as we go to larger radii, the relative importance of the old episode 
increases.
This produces a positive age radial gradient and a negative metallicity radial gradient, which
are steeper between the intermediate and outer region.
The gradient in the SFH of Carina is consistent with what had already been reported by
various previous studies \citep[e.g.,][]{battaglia12,mcMonigal14,deBoer14}.

One consequence of the spatial variation of Carina's SFH is the fact that the
intermediate-age population is practically negligible at large radii.
The lack of intermediate-age stars in the outer region of Carina
suggests that, once this population started forming, tidal forces from the Milky Way, or any
other physical process\footnote{For example relaxation can not be responsible for the 
spatial distribution of the different stellar populations in Carina, since the relaxation times
in dwarf galaxies are much larger than the age of the Universe \citep[e.g.,][]{freeman08} }, 
were not strong enough to move a significant fraction of these stars to the outer region.
Otherwise, a large number of intermediate-age stars would be present in the outskirts of
Carina, and thus, we would not observe such a strong break in the SFH and MDF of Carina
between the intermediate and outer regions.
Since the radial trends in Carina do not seem to be governed by the tidal influence of the
Milky Way, we suggest that this hints that the SFH is not dominated by tidal forces either.
In this line, the fact that we find two episodes of star formation in Carina instead of three, as 
it has often been claimed \citep{hurley-keller98,rizzi03,pilkington12,mcMonigal14},
makes it harder to explain the star formation episodes in Carina as the 
result of periodic close passages to the Milky Way \citep[as done in][]{pasetto11}, 
even more so 
since proper motions suggest that the orbital period of Carina is close to
$2$\,Gyr \citep{piatek03,pasetto11}.

Another mechanism proposed to explain the SFH of Carina is gas accretion from the Milky 
Way \citep[e.g.,][]{lemasle12,deBoer14}.
In this view, the first episode would have ended due to gas depletion 
and subsequent star formation would have been stopped
for a 
few gigayears 
until the dwarf galaxy renewed its gas content thanks to
an inflow from the Milky Way.
Even though this is a 
plausible scenario, the in-falling gas should have had the same 
metallicity as the stars from the end of the first episode to reproduce the age-metallicity
relation in Carina.
A simpler way to explain why the metallicity of the stars at the end of the first episode 
coincides with 
that of the stars at the beginning of the second episode
is to assume that the gas that formed the second episode in Carina was 
enriched within the same galaxy by the first generation of stars.

Internal evolution would also be consistent with the 
mean metallicities and ages at 
different Carina regions, which according to what we see in Figure~\ref{3D_SFHs} are 
roughly constant throughout the extent of the galaxy.
This 
would not necessarily be the case if the gas that formed those stars 
had been accreted at different moments and each region had different levels of
influence from each inflow.
 
For all these reasons, we conclude that the SFH of Carina must be dominated by internal 
evolution.
Our interpretation is that after the first episode formed the majority of the stars in 
Carina, the physical conditions of the gas made the galaxy unable to
form stars for a few gigayears, until it contracted to activate star formation again.

We claim that the temporal gap in star formation in Carina might have been produced by
feedback processes. It has been often claimed \citep[e.g.,][]{kaviraj07} that feedback can
significantly decrease the efficiency of star formation.
In the large halo mass regime (galaxy clusters) this would be due to active galactic nuclei.
For low halo masses (dwarf galaxies) this process have been proposed to be caused by 
supernovae \citep[e.g.,][]{maclow89}, and more recently, stellar-mass black holes \citep[e.g.]
[]{leigh13}.

For example, it has been proposed by simulations \citep[e.g.,][]{revaz09}, that the star 
formation of systems with low initial mass ($M_{\rm{i}}<3\times10^{8} M_{\rm{\odot}}$), 
could be self-regulated.
In these systems, star formation is followed by shocks of energy released by the 
supernovae that heat and expand the gas.
Given that the gas density is relatively low, the gas cooling times are large and can take up 
to several gigayears to contract to form stars again.
If this is the case, the gas available decreases, due to the first episode of star formation,
and hence it should contract to a radius smaller than the original to reach the density 
conditions necessary to reactivate the star formation.
This coincides with what we observe in this study, where the second episode occurred 
preferentially in a 
region contained within
approximately $0.6\times r_{\rm{tidal}}$ of Carina.
Unlike the 
majority of studies about Carina's SFH, we found that the first episode of 
star formation was the most important one.
We also know that the second star formation episode occurred during a longer period of time.
This could explain why, unlike the first episode of star formation, the second one did not
halt the star formation and the galaxy could keep forming stars more or less regularly from
$\sim8$ to $\sim2$ gigayears ago.

But this scenario prompts the question of what is different in Carina to make it have a qualitatively different SFH than any other dSph galaxy?
There seems to be a trend in the Local Group wherein more luminous 
galaxies (Irregular galaxies and brightest classical dwarfs) display more complex SFHs with 
important components from intermediate-age and young star formation; while dimmer galaxies
(dimmest classical dwarfs and ultra faints) are composed of single old episodes of star
formation \citep[e.g.,][]{tolstoy09,brown14}.
Galaxies with larger masses have larger potential wells and their gas density is generally 
larger. Given that the cooling of the gas is directly proportional to its density \citep[see 
details for example on][]{revaz09}, these galaxies have shorter cooling times.
This enables them to have significant star formation throughout all their history.
On the other hand, galaxies with lower masses cannot retain the gas to form stars after the 
first episode or the gas cannot cool and contract again.
Carina has a luminosity similar to the dimmest classical dwarf galaxies, but unlike them,
it has an important component of intermediate-age stars.
However, there is evidence that this galaxy may have lost a significant fraction of its mass in the 
past \citep{munoz06b,majewski00b,munoz08a} and this process is ongoing.
This means that in the past Carina could have been significantly brighter than classical 
dwarfs with currently similar luminosities.
Thus, its episodic SFH could represent an intermediate regime between the
massive dwarfs with continuous star formation and the least massive dwarfs with 
only one old episode of star formation.
In this context, Carina could have been massive enough to retain the gas after the first 
episode of star formation, but not massive enough to avoid significant expansion of the gas 
which stopped the star formation, and given the relatively low gas densities, it could only 
cool down and contract to form stars again after a couple of gigayears.

\section{Conclusions}

In this work we have presented the spatially resolved star formation history (measured
in the inner, middle, and outer region) of the Carina dwarf spheroidal galaxy using deep,
wide-field $g$, $r$ imaging and the metallicity distribution function from the data of
\citet{koch06}.
This is the first time a combination of depth and coverage of this quality is used to
derive the SFH of Carina, enabling us to trace its different populations with unprecedented 
accuracy. The main results of this work can be summarized as followed:

1. The SFH of Carina shows a majority of old metal poor stars accounting for 
$54\pm4\%$ of the stellar mass and an intermediate-age population with increasing 
metallicity with time.
The fraction of old stars could be larger considering the possibility that tidal
influence from the Milky Way might have removed preferentially old stars from Carina,
given that these stars are distributed in a more extended region.

2.  Both episodes are separated by a period of no star formation. This temporal gap started
$\sim10\,$Gyr ago and stopped $\sim8\,$Gyr ago.

3. Carina displays a positive age radial gradient and a negative metallicity radial gradient.
The inner region is dominated by the intermediate-age population, the middle region is
composed of a similar fraction of old and intermediate-age stars, and the outer region is
dominated by the old metal poor population, with an almost negligible component of
intermediate-age stars.

4. Results are consistent with a total stellar mass of
$1.60\pm0.09\times 10^{6} \, M_{\rm{\odot}}$ within 
$1.3\times r_{\rm{tidal}}$, and $1.45\pm0.12\times 10^{6} \, M_{\rm{\odot}}$ within
the nominal $r_{\rm{tidal}}$. The latter value is larger than previous values derived for 
Carina. For example, \citet{deBoer14} derived a value of $1.07\pm0.08\times 10^{6} \, 
M_{\rm{\odot}}$ within the nominal $r_{\rm{tidal}}$. We attribute the difference in the
masses derived to the increased depth of our photometry that enabled us to
detect more faint old main-sequence stars.

5. The star formation is consistent with no young ($<2$\,Gyr) stars. 
We calculated the fraction of blue plume stars in Carina and the fraction of blue stragglers
in $4$ other local dwarf spheroidal galaxies.
These fractions were calculated using the same mass range (in terms of the mass of the
main-sequence turnoff) for both cases, in order to make a meaningful comparison between
both fractions.
Given that blue straggler fraction in the local dwarf galaxies is constant and that the
blue straggler fraction of Carina is completely consistent with this value, 
we concluded that the blue plume in this galaxy is consistent with being
composed of blue stragglers.

6. The spatially resolved SFH is consistent with being dominated by internal evolution as 
opposed to tidal influence from the Milky Way.
Our hypothesis is that Carina formed the majority of its stars in a first episode of
star formation,  then the star formation
ceased for a couple of gigayears due to an internal process (e.g., gas heating) and 
finally, after gas cooling, re-accreted the gas producing the second episode of star formation.

\acknowledgments

F.A.S.~acknowledges support from CONICYT Anillo project ACT-1122
R.R.M.~acknowledges support from CONICYT through project BASAL PFB-06 and from the 
FONDECYT project N$^{\circ}1120013$. 
M.~G.~acknowledges support from the National Science Foundation under award 
number AST-0908752 and the Alfred P.~Sloan Foundation.
S.G.D.\ was supported in part by the NSF grants AST-1313422, AST-1413600, and 
AST-1518308. A.E.G.~acknowledges support from FONDECYT grant 3150570.

\appendix

\section{Determining Blue Plume and Blue Straggler Fractions}

To calculate the contribution of blue stragglers over young stars in the blue plume of Carina
we compare the fraction of blue plume stars in this galaxy with the fraction of blue 
stragglers in other $4$ local dSph galaxies: Sculptor, Ursa Minor, Draco and Sextans.
We will denominate these the \emph{comparison dwarf galaxies}, and we use them
as a control sample.
Since the blue straggler fraction in local dwarf spheroidals has been found to be constant
\citep{momany07a,santana13a}, the difference between the blue plume fraction in Carina 
and the blue straggler fraction in the comparison local dwarf galaxies can be used to
estimate the fraction of genuine young stars in Carina, given that there are no young stars
in the comparison dwarf galaxies \citep[e.g.,][and references therein]{tolstoy09}.
To compare the contribution of blue stragglers in Carina with the one in the comparison
dwarf galaxies, six important points have to be considered in order to make a
consistent selection criteria.\\
1) Carina is composed of $2$ main stellar populations, one old and one at intermediate
age. Thus, there are blue stragglers in Carina originating from each of these $2$ stellar
populations, whose numbers we denominate $BSS_{\rm OLD}$ and $BSS_{\rm INT}$,
respectively.
\\
2) A large fraction of $BSS_{\rm OLD}$ in Carina cannot be detected in the CMD because
their location
overlaps with the much more numerous intermediate-age main-sequence stars.
Therefore, from the CMD we cannot measure $BSS_{\rm OLD}$, and thus, for Carina the
objective is to calculate $BSS_{\rm INT}$.
\\
3) The main-sequence turnoff of the intermediate-age population of Carina is located
close to the horizontal branch stars.
Thus, we have to restrict the CMD region were
we count the blue plume stars in Carina in order to avoid extremely large contamination in
our measurements.
Accordingly, the blue plume box defined for Carina includes only the blue stragglers with
masses between $1.4$ to $1.55$ times the mass of the main-sequence turnoff of the
intermediate-age population.
We denominate this quantity as $BSS_{\rm INT,1.55}$.
\\
4) The blue plume box of Carina so defined is located far from the old main-sequence
turnoff, but it still includes a small fraction of BSS originated from the old population.
Therefore, to estimate $BSS_{\rm INT,1.55}$ in Carina, we have to subtract this small 
contribution.
\\
5) The blue plume and blue straggler numbers measured for all these galaxies have to be
normalized to the total number of stars of the stellar population from which the blue
stragglers were formed.
In this way, we can compare the blue plume star fraction in Carina with the fraction
of blue stragglers in the comparison local dwarf galaxies.
Because we are using the blue plume box of Carina to determine $BSS_{\rm INT}$ (plus the
possible contribution of young stars) we normalize the number of blue plume stars
dividing it by the number of stars from the intermediate-age population.
For the comparison dwarf galaxies, their CMDs are consistent with being
composed of stars which are old ($\sim12$\,Gyr) and metal poor.
Therefore, the blue stragglers in these systems originate from the old (and only) stellar 
population.
\\
6) All the type of stars we consider for this analysis (blue plume, blue stragglers and
total stars) are counted in a region that goes from the center until 2 times
its half light radius, which corresponds to the spatial coverage of our photometry.
This volume is representative of the entire galaxy, because in all cases includes the vast
majority of the stars.
Nevertheless, some fraction of blue plume or blue straggler stars are not detected simply
because they are located outside our coverage.
We normalize the populations dividing by the number of stars located inside 2 times the
half light radius instead of the \emph{total} number of stars or the total luminosity of the
system.
\\
Taking into account these six considerations, we calculate the BSS fraction for the 
comparison dwarf galaxies as
\begin{equation}
F_{\rm BSS,dwarfs}\equiv BSS_{\rm TOT}/N_{\rm TOT}=BSS_{\rm OLD}/N_{\rm OLD}\quad.
\label{bss_frac_dwarfs}
\end{equation}
Because these galaxies are old, there are no intermediate-age blue stragglers in their CMDs. 
Therefore, they can be counted in the entire magnitude range covering right up from the
main-sequence turnoff until just below the horizontal branch.
For this reason, $BSS_{\rm TOT}$ can be measured directly for the comparison dwarf 
galaxies.
On the other hand, $N_{\rm TOT}$ can not be measured directly because a large fraction
of the stars are not detected given that their magnitudes are fainter than our detection limit.
Therefore, we used the number of RGB stars in these systems as tracers of the total number 
of stars.
We chose them as tracers because they are well above our detection
limit and they are not affected by completeness.
Moreover, they are very numerous compared to bright stars in other stages
of stellar evolution, such as horizontal branch, or asymptotic giant branch stars.
We note that \citet{santana13a}, using
a sample of globular clusters and local dwarf galaxies, demonstrated that the number of
RGB stars grows linearly with the luminosity of the system, indicating that there is a simple 
linear transformation between the number of RGB stars and the total number of stars.

\begin{figure*}
\includegraphics[width=\textwidth]{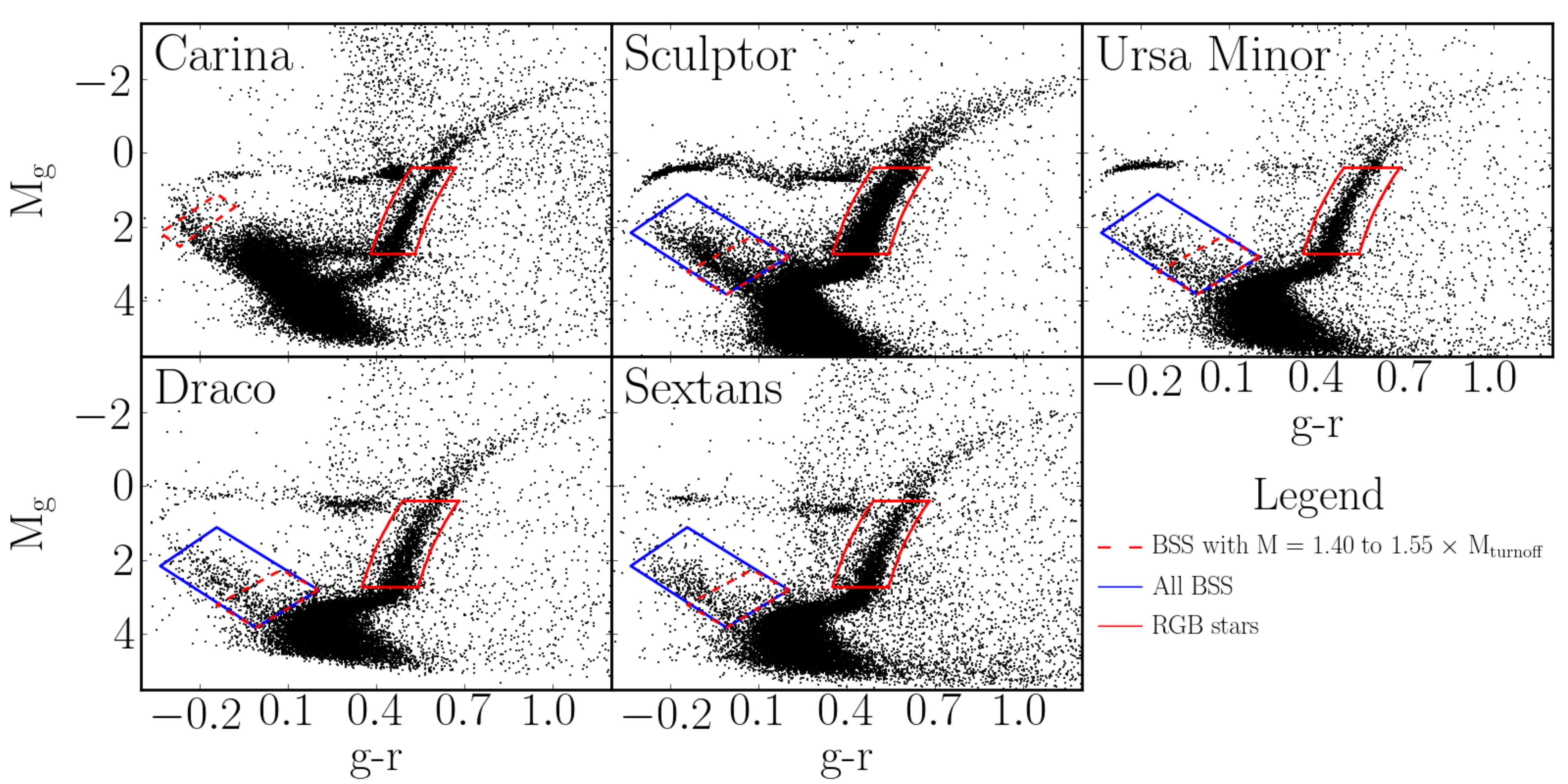}
\caption{CMDs showing the boxes used to count blue plume stars in Carina blue stragglers 
in the comparison dwarf galaxies and RGB stars in all the galaxies.
Red solid boxes show the region where we counted the RGB stars.
The red dashed box in the CMD of Carina indicates the blue plume box, which was
defined to include stars from $1.40$ to $1.55$ times 
the mass of the main sequence turnoff of the intermediate-age population.
The red dashed box in the comparison dwarf galaxies indicates one of the
blue straggler boxes used for these galaxies, which was defined to include stars from
$1.40$ to $1.55$ times the mass of the main sequence turnoff of the old stellar population.
The solid blue boxes indicate the second blue straggler box used for the comparison
dwarf galaxies, which was defined to count all the BSS in these systems.
}
\label{BSS_cmds}
\end{figure*}

Therefore, we calculate $N_{\rm TOT}$ as:
\begin{equation}
N_{\rm TOT}=RGB_{\rm TOT}/X_{\rm RGB,OLD}\quad,
\label{n_tot_def}
\end{equation}
in which $RGB_{\rm TOT}$ is the total number of RGB stars detected
inside the red solid boxes shown in Figure \ref{BSS_cmds}.
These boxes are defined around an isochrone corresponding to an age of $12$\,Gyr and an
abundance of [Fe/H] of $-1.9$, which is a good representation of the stellar populations
of all these galaxies. The width of the RGB box is $0.095$ magnitudes at each side of the
isochrone, including a magnitude range that goes from $4.9$
to $2.4$ magnitudes below the RGB tip. 
$X_{RGB,OLD}$ corresponds to the fraction that selected RGB stars represent from the entire
number of stars of an old (age=$12$\,Gyr) metal-poor ([Fe/H]=$-1.9$) population.
To determine this value we calculate the mass range of the stars in the RGB box, and then
used a Kroupa IMF \citep{kroupa01} to translate that mass range into a fraction of the 
stars originally formed.
The value obtained for $X_{RGB,OLD}$ is $(5.26\pm0.34)\times 10^{-4}$.
We calculated the uncertainty by repeating the process with different isochrones of 
slightly different ages and metallicities consistent with the intermediate-age population of 
the comparison dwarf galaxies, and taking the uncertainty as the standard deviation
of the different values obtained for $X_{RGB,OLD}$.
Replacing Equation \ref{n_tot_def} into Equation
\ref{bss_frac_dwarfs} we obtain
\begin{equation}
F_{\rm BSS,dwarfs}=X_{RGB,OLD} \times \frac{BSS_{\rm TOT}}{RGB_{\rm TOT}}\quad,
\label{bss_frac_dwarfs_solved}
\end{equation}
whose values are given in the seventh column of Table~\ref{star_counts_cdg}.

For the case of Carina, the blue plume includes,
in principle, both blue stragglers and young stars.
Therefore, the number of blue plume stars measured in Carina is given by
\begin{equation}
N_{\rm BP,Car}=BSS_{\rm INT,BOX}+BSS_{\rm OLD,BOX}+YS_{\rm BOX}\quad,
\label{bp_number_carina1}
\end{equation}
where the BOX subscript refers to the CMD location indicated in Figure~\ref{BSS_cmds}
(red dashed box).
In Equation \ref{bp_number_carina1}, $YS_{\rm BOX}$ represents the number of young stars 
falling in the blue plume box.
Based on the magnitude limits of the blue plume box defined for Carina,  we calculate that 
the blue stragglers from the intermediate-age population included in the box have
masses ranging from $1.40$ to $1.55$ times the mass of the intermediate-age population
main-sequence turnoff of Carina.
We estimate this mass range translating the magnitude limits of the blue plume box of
Carina into stellar masses based on a very young (age=1\,Gy) isochrone with the 
metallicity of the intermediate-age population of Carina ([Fe/H]=$1.4$).
Given that $BSS_{\rm INT,BOX}$ only includes the stars in that mass range we will 
denominate this quantity as $BSS_{\rm INT,1.55}$.
Analogously, $BSS_{\rm OLD,BOX}$ only includes the brightest blue stragglers that come
from the old population of Carina, and thus, we denominate this quantity as
$BSS_{\rm OLD,BR}$. Therefore, Equation~\ref{bp_number_carina1} can be re-written as
\begin{equation}
N_{\rm BP,Car}=BSS_{\rm INT,1.55}+BSS_{\rm OLD,BR}+YS_{\rm BOX}\quad.
\label{bp_number_carina2}
\end{equation}

The final goal of this analysis is to compare the blue plume fraction of Carina with the blue 
straggler fraction of the comparison dwarf galaxies.
All the stars in the comparison dwarf galaxies are old and metal poor, and 
unlike Carina, their blue stragglers come from a \emph{single} stellar population.
For this reason, we subtract the contribution from the old blue stragglers to the 
blue plume counts in Carina.
In this way, we can compare the blue stragglers from a single populations in Carina with
those of the comparison dwarf galaxies.
Estimating $BSS_{\rm OLD,BR}$ entails calculating the fraction of blue stragglers from the 
old population falling in the blue plume box defined for Carina.
We denominate this quantity as $X_{\rm BR,TOT}$ and is defined as
\begin{equation}
X_{\rm BR,TOT}\equiv\frac{BSS_{\rm OLD,BR}}{BSS_{\rm OLD}}\quad.
\label{r_bright_tot_def}
\end{equation}
To estimate this quantity  we can use the measurements directly from  
Table~\ref{star_counts_cdg}.
Given that $X_{\rm BR,TOT}$ is fairly constant in the comparison dwarf galaxies, for
Carina we estimate it as the average of the values found for the comparison dwarf 
galaxies and its uncertainty as the standard deviation. In this way, we obtain
\begin{equation}
X_{\rm BR,TOT}=0.076\pm0.020\quad,
\label{r_bright_tot_value}
\end{equation}
which implies
\begin{equation}
BSS_{\rm OLD,BR}=X_{\rm BR,TOT}\times \frac{BSS_{\rm OLD}}{N_{\rm OLD}}\times
N_{\rm OLD}
\label{bss_old_bright2}
\end{equation}
Table~\ref{star_counts_cdg} also shows the fractions of old blue stragglers over the total 
number of old stars for the comparison dwarf galaxies.
This value is almost constant in these galaxies, with a standard deviation that
represent only $\sim 7\%$ of the average.
Moreover, \citet{santana13a} found that the fraction of blue stragglers in local
dwarf galaxies is practically constant.
For these reasons, we estimate $\frac{BSS_{\rm OLD}}{N_{\rm OLD}}$ for Carina
as the average value found for dwarf galaxies with the uncertainty equal to  the standard
deviation, that is,
\begin{equation}
\frac{BSS_{\rm OLD}}{N_{\rm OLD}}=(1.78\pm0.13)\times 10^{-4}\quad,
\label{f_bss_old}
\end{equation}
which implies that (see Equation \ref{bss_old_bright2})
\begin{equation}
BSS_{\rm OLD,BR}=(1.36\pm0.19)\times 10^{-5}\times N_{\rm OLD}\quad.
\label{bss_old_bright3}
\end{equation}
To calculate
$N_{\rm OLD}$, we estimate it in an analogous way as $N_{\rm TOT}$ for the 
comparison dwarf galaxies, that is, by using the RGB stars as tracers of the total number of 
stars formed in the corresponding star formation episode.
The RGB box defined for Carina (red continuous box of 
Figure~\ref{BSS_cmds}) includes RGB stars that belong to the old population
($RGB_{\rm OLD}$) and RGB stars that belong to the intermediate-age population
($RGB_{\rm INT}$).
By using the results obtained from the SFH analysis performed on Carina, we calculate that
$42\%$ of the stars in the RGB box of Carina belong to the old population.
Additionally, as we pointed out earlier, $X_{\rm RGB,OLD}$=$(5.26\pm0.34)\times 10^{-4}$, thus
 we obtain that
\begin{equation}
N_{\rm OLD}=0.42 \frac{\times RGB_{\rm TOT}}{X_{\rm RGB,OLD}}
=(799\pm52)\times RGB_{\rm TOT}\quad,
\label{n_old_value}
\end{equation}
and using Equation~\ref{bss_old_bright3}, we obtain that
\begin{equation}
BSS_{\rm OLD,BR}=(0.011\pm0.002)\times  RGB_{\rm TOT}\quad.
\label{bss_old_bright4}
\end{equation}
By subtracting this equation from
 Equation~\ref{bp_number_carina2} and reordering we get that
\begin{equation}
N_{\rm BP,Car}-0.011\times  RGB_{\rm TOT}=BSS_{\rm INT,1.55}\left( 1+
\frac{YS_{\rm BOX}}{BSS_{\rm INT,1.55}} \right)\quad.
\label{bp_number_carina4}
\end{equation}
The next step is to estimate the total number of blue stragglers in Carina that come from
the intermediate-age population ($BSS_{\rm INT}$) based on the value of
$BSS_{\rm INT,1.55}$.
For this 
we will assume that the fraction of blue stragglers in the mass range of
1.4 to 1.55 times the mass of the main sequence turnoff over the total number of blue 
stragglers of that particular population is the same for the old and
intermediate-age populations. That is,
\begin{equation}
\frac{BSS_{\rm INT,1.55}}{BSS_{\rm INT}}=\frac{BSS_{\rm OLD,1.55}}{BSS_{\rm OLD}}\quad.
\label{bss_assumption}
\end{equation}
This assumption is based on $2$ observations. The first one is that there is no significant
dependence of the fraction of blue stragglers on the age of the stellar population for ages
larger than $3$--$4$\,Gyr \citep{deMarchi06,ahumada07,xin07}.
The second one is that the fraction
$\frac{BSS_{\rm OLD,1.55}}{BSS_{\rm OLD}}$ is practically constant among the comparison
dwarf galaxies, displaying a standard deviation that represents only $7\%$ of the average
value.
By calculating $\frac{BSS_{\rm OLD,1.55}}{BSS_{\rm OLD}}$ for the comparison dwarf 
galaxies we obtain
\begin{equation}
\frac{BSS_{\rm OLD,1.55}}{BSS_{\rm OLD}}=0.528\pm0.037\quad,
\label{bss_mass_frac_old}
\end{equation}
which implies
\begin{equation}
BSS_{\rm INT,1.55}=0.528\times BSS_{\rm INT}\quad.
\label{bss_int_carina_value}
\end{equation}
Therefore 
\begin{equation}
1.894 ( N_{\rm BP,Car}-0.011\times  RGB_{\rm TOT})=BSS_{\rm INT}
\left( 1+\frac{YS_{\rm BOX}}{BSS_{\rm INT,1.55}} \right)\quad.
\label{bp_number_carina6}
\end{equation}
Now to calculate the blue plume \emph{fraction} of Carina, $F_{\rm BP,Car}$, we have to
divide Equation~\ref{bp_number_carina6} by the total number of stars from the
intermediate-age population, $N_{\rm INT}$.
By doing this, we obtain
\begin{equation}
F_{\rm BP,Car}=
\frac{1.894 ( N_{\rm BP,Car}-0.011\times  RGB_{\rm TOT})}{N_{\rm INT}}=
F_{\rm BSS,Car}\left( 1+\frac{YS_{\rm BOX}}{BSS_{\rm INT,1.55}} \right)\quad,
\label{bp_number_carina7}
\end{equation}
where F$_{\rm BSS,Car}$ is the fraction of genuine blue stragglers in Carina, the quantity
that has been consistently calculated to be compared to the fraction of blue stragglers
in the comparison dwarf galaxies ($F_{\rm BSS,dwarf}$).
Now the only thing remaining is to calculate $N_{\rm INT}$, which we will do in the same 
way as for $N_{\rm OLD},$ by using the number of RGB stars as a tracer of the total number 
of stars.
Analogously as we defined $N_{\rm OLD}$ in Equation~\ref{n_tot_def}, we define
$N_{\rm INT}$ as
\begin{equation}
N_{\rm INT}=RGB_{\rm INT}/X_{\rm RGB,INT}\quad,
\label{n_int_def}
\end{equation}
where $X_{\rm RGB,OLD}$ corresponds to the fraction $RGB_{\rm INT}$ 
from the entire intermediate-age population.
To model the intermediate-age population of Carina, we will use a
Dartmouth \citep{dotter08} isochrone with an age 6 Gyr and an abundance of [Fe/H]=-1.4,
which produces an accurate fit to the CMD location of the intermediate-age population of
Carina.
As we did for $X_{\rm RGB,OLD}$, we determine $X_{\rm RGB,INT}$ by translating the 
magnitude limits of the RGB box into a stellar mass range, according to the isochrone 
representative of the population. 
Then, using a Kroupa IMF \citep{kroupa01} we translate that mass range into a fraction of 
the stars originally formed for the intermediate-age population.
By doing this, we obtain a value of $(8.68\pm0.39)\times 10^{−4}$ for $X_{\rm RGB,INT}$.
By using the results obtained from the SFH analysis performed on Carina, we 
calculate that 58\% of the stars in the RGB box of Carina belong to the
intermediate-age population.
Thus, by replacing these values into Equation~\ref{n_int_def}, we obtain that
\begin{equation}
N_{\rm INT}=
\frac{0.58\times RGB_{\rm TOT}}{8.68\times 10^{-4}}=
668 \times RGB_{\rm TOT}\quad.
\label{n_int_value}
\end{equation}
Then, by replacing this equation into Equation~\ref{bp_number_carina7} we can finally
obtain
\begin{equation}
F_{\rm BP,Car}=
2.84 \times 10^{-3} \times \frac{N_{\rm BP,Carina}}
{N_{\rm RGB}}-3.08\times 10^{-5}=
F_{\rm BSS,Car}\left( 1+\frac{YS_{\rm BOX}}{BSS_{\rm INT,1.55}} \right)\quad.
\label{bp_number_carina8}
\end{equation}
 Which allows us to calculate $F_{\rm BP,Car}$ directly from the observables
$N_{\rm BP,Car}$ and $N_{\rm RGB}$.
These values correspond respectively to the number of stars in the dashed and solid red 
line on Carina's CMD in Figure~\ref{BSS_cmds}.
The right side of the equation depends on the fraction 
$\frac{YS_{\rm BOX}}{BSS_{\rm INT,1.55}}$.
This quantity is the ratio of genuine young stars over the genuine blue stragglers that are 
located in the blue plume box of Carina, and hence, it can significantly help us elucidating 
the nature of the stars in this CMD region.
The blue plume fraction of Carina is then the genuine blue straggler fraction multiplied
by a correction factor that depends only on the fraction of young stars in the blue plume
box.
By replacing the observed number of stars in the different CMD boxes (listed in 
tables~\ref{star_counts_cdg} and \ref{star_counts_carina} ) in 
Equations~\ref{bss_frac_dwarfs_solved} and
\ref{bp_number_carina8} we obtain the values for the blue plume fraction for Carina
($F_{\rm BP,Car}$), and the average blue straggler fraction for the comparison dwarf 
galaxies ($\overline{F_{\rm BSS,dwarfs}}$).
\begin{equation}
F_{\rm BP,Car}=
F_{\rm BSS,Car}\left( 1+\frac{YS_{\rm BOX}}{BSS_{\rm INT,1.55}} \right)=
(1.89\pm0.25)\times 10^{-4}\quad,
\label{bp_number_carina9}
\end{equation}
and
\begin{equation}
\overline{F_{\rm BSS,dwarfs}}=(1.79\pm0.13)\times 10^{-4}\quad,
\label{bss_frac_dwarfs_value}
\end{equation}
where the error cited for $\overline{F_{\rm BSS,dwarfs}}$ corresponds to the standard 
deviation calculated for the comparison dwarf galaxies.
If we compare these last two equations we can see that the blue plume fraction of Carina
is consistent within the errors with the blue straggler fraction in the comparison dwarf
galaxies, or equivalently,
$YS_{\rm BOX}$
is consistent with being equal to zero.
Based on this result, we conclude that there are no genuine young (age$<$2\,Gyr) stars
in Carina, and that the stars in the blue plume of this galaxy are instead blue stragglers.
This is illustrated  in Figure~\ref{BSS_trend}, where we plot the blue
straggler fraction of all the comparison dwarf galaxies and the blue plume fraction of
Carina.
Based on our calculations, we derive that the maximum fraction of young
stars in the blue plume of Carina consistent with our uncertainties is $16.5\%$.

According to our original SFH derivation for Carina, we calculated that
$(1.45\pm0.19)\%$ of the stars in Carina are younger than 2\,Gyr.
Thus, if $16.5\%$ of those young stars were authentic young stars as opposed to
blue stragglers, then, only the $0.24\%$ of the stars would be younger than 2\,Gyr.
This fraction is practically negligible, and given that the highest young star fraction
that is consistent with our measurements is comparable to the error in the SFH derivation,
we confirm our previous conclusion that there are no authentic young stars in Carina,
and that its blue plume is composed of blue stragglers.

%
%
%

%

\floattable%
\begin{deluxetable}{ccccccccc}
\rotate
\tablecolumns{9}
\tablewidth{0pt}
\tabletypesize{\footnotesize}
\tablecaption{Stellar Counts for the Comparison Dwarf Galaxies}
\tablehead{
\colhead{Object} & 
\colhead{$BSS_{\rm{OLD}}$\tablenotemark{1}} & 
\colhead{$BSS_{\rm{OLD,1.55}}$\tablenotemark{2}} &
\colhead{$BSS_{\rm{OLD,BR}}$\tablenotemark{3}} & 
\colhead{$RGB_{\rm{TOT}}$}& 
\colhead{$N_{\rm OLD}$\tablenotemark{4}} &
\colhead{$F_{\rm BSS}$\tablenotemark{5}} &
\colhead{$\frac{BSS_{\rm{OLD,BR}}}{BSS_{\rm{OLD}}}$} &
\colhead{$\frac{BSS_{\rm{OLD,1.55}}}{BSS_{\rm{OLD}}}$}
}
\startdata
Sculptor&1434&768&80&4746&9.03$\times10^{6}$&  $(1.59\pm0.10)\times10^{-4}$  &0.055&0.534\\
Ursa Minor&567&310&37&1584&3.01$\times10^{6}$&  $(1.88\pm0.12)\times10^{-4}$  &0.066&0.547\\
Draco&595&330&59&1697&3.23$\times10^{6}$&  $(1.84\pm0.12)\times10^{-4}$  &0.100&0.554\\
Sextans&545&258&45&1564&2.98$\times10^{6}$&  $(1.83\pm0.12)\times10^{-4}$  &0.083&0.474\\
\hline
Average&&&&&& $1.79\times 10^{-4}$ &0.076&0.528\\
Standard Deviation&&&&&& $1.34\times 10^{-5}$ &0.020&0.037\\
\enddata
\tablenotetext{1}{BSS from the old population}
\tablenotetext{2}{BSS from the old population with  M=($1.4$--$1.55$)$\times M_{\rm 
turnoff}$}
\tablenotetext{3}{BSS from the old population bright enough to fall in the blue plume box defined for 
Carina}
\tablenotetext{4}{Number of stars from the old stellar population}
\tablenotetext{5}{Fraction of blue stragglers as measured from Equation~\ref{bss_frac_dwarfs}}
\label{star_counts_cdg}
\end{deluxetable}

\floattable%
\begin{deluxetable}{ccccccccccc}
\rotate
\tablecolumns{11}
\tablewidth{0pt}
\tabletypesize{\scriptsize}
\tablecaption{Stellar Counts for Carina}
\tablehead{ 
\colhead{$RGB_{\rm{TOT}}$\tablenotemark{1}} & 
\colhead{$RGB_{\rm{INT}}$\tablenotemark{2}} & 
\colhead{$RGB_{\rm{OLD}}$\tablenotemark{3}} & 
\colhead{$N_{\rm{INT}}$\tablenotemark{4}} & 
\colhead{$N_{\rm{OLD}}$\tablenotemark{5}} & 
\colhead{$BSS_{\rm{OLD}}$\tablenotemark{6}} & 
\colhead{$BSS_{\rm{OLD,BR}}$\tablenotemark{7}} &
\colhead{$BSS_{\rm{TOT,1.55}}$\tablenotemark{8}} & 
\colhead{$BSS_{\rm{INT,1.55}}$\tablenotemark{9}} & 
\colhead{$BSS_{\rm{INT}}$\tablenotemark{10}} &  
\colhead{$F_{\rm BP}$\tablenotemark{11}} 
}
\startdata
1724&1000&725&  $ 1.15\times10^{6} $  &  $ 1.38\times10^{6} $  & 246 & 19 & 134 & 115 & 218 &
$(1.89\pm0.25)\times 10^{-4}$
\\
\enddata
\tablenotetext{1}{Total number of stars falling in the RGB box defined for Carina}
\tablenotetext{2}{RGB stars from the intermediate-age population}
\tablenotetext{3}{RGB stars from the old population}
\tablenotetext{4}{Number of stars from the old stellar population}
\tablenotetext{5}{Number of stars from the intermediate-age population}
\tablenotetext{6}{BSS from the old population}
\tablenotetext{7}{BSS from the old population bright enough to fall in the blue plume box}
\tablenotetext{8}{BSS that fall in the blue plume box of Carina}
\tablenotetext{9}{BSS form the intermediate-age population with M=($1.4$--$1.55$)$\times M_{\rm turnoff}$}
\tablenotetext{10}{BSS from the intermediate-age population}
\tablenotetext{11}{Fraction of blue stragglers as measured from Equation~\ref{bp_number_carina8}}
\label{star_counts_carina}
\end{deluxetable}

\clearpage\bibliography{manuscript}

\begin{thebibliography}{67}
\expandafter\ifx\csname natexlab\endcsname\relax\def\natexlab#1{#1}\fi

\bibitem[{{Ahumada} \& {Lapasset}(2007)}]{ahumada07}
{Ahumada}, J.~A., \& {Lapasset}, E. 2007, \aap, 463, 789

\bibitem[{{Aparicio} \& {Hidalgo}(2009)}]{aparicio09}
{Aparicio}, A., \& {Hidalgo}, S.~L. 2009, \aj, 138, 558

\bibitem[{{Armandroff} \& {Da Costa}(1991)}]{armandroff91}
{Armandroff}, T.~E., \& {Da Costa}, G.~S. 1991, \aj, 101, 1329

\bibitem[{{Battaglia} {et~al.}(2012){Battaglia}, {Irwin}, {Tolstoy}, {de Boer},
  \& {Mateo}}]{battaglia12}
{Battaglia}, G., {Irwin}, M., {Tolstoy}, E., {de Boer}, T., \& {Mateo}, M.
  2012, \apjl, 761, L31

\bibitem[{{Bono} {et~al.}(2010){Bono}, {Stetson}, {Walker}, {Monelli},
  {Fabrizio}, {Pietrinferni}, {Brocato}, {Buonanno}, {Caputo}, {Cassisi},
  {Castellani}, {Cignoni}, {Corsi}, {Dall'Ora}, {Degl'Innocenti}, {Fran{\c
  c}ois}, {Ferraro}, {Iannicola}, {Nonino}, {Moroni}, {Pulone}, {Smith}, \&
  {Thevenin}}]{bono10}
{Bono}, G., {et~al.} 2010, \pasp, 122, 651

\bibitem[{{Brown} {et~al.}(2014){Brown}, {Tumlinson}, {Geha}, {Simon},
  {Vargas}, {VandenBerg}, {Kirby}, {Kalirai}, {Avila}, {Gennaro}, {Ferguson},
  {Mu{\~n}oz}, {Guhathakurta}, \& {Renzini}}]{brown14}
{Brown}, T.~M., {et~al.} 2014, \apj, 796, 91

\bibitem[{{Cannon} {et~al.}(1977){Cannon}, {Hawarden}, \& {Tritton}}]{cannon77}
{Cannon}, R.~D., {Hawarden}, T.~G., \& {Tritton}, S.~B. 1977, \mnras, 180, 81P

\bibitem[{{Carretta} \& {Gratton}(1997)}]{carretta97}
{Carretta}, E., \& {Gratton}, R.~G. 1997, \aaps, 121, 95

\bibitem[{{Coppola} {et~al.}(2013){Coppola}, {Stetson}, {Marconi}, {Bono},
  {Ripepi}, {Fabrizio}, {Dall'Ora}, {Musella}, {Buonanno}, {Ferraro},
  {Fiorentino}, {Iannicola}, {Monelli}, {Nonino}, {Pulone}, {Th{\'e}venin}, \&
  {Walker}}]{coppola13}
{Coppola}, G., {et~al.} 2013, \apj, 775, 6

\bibitem[{{de Boer} {et~al.}(2014){de Boer}, {Tolstoy}, {Lemasle}, {Saha},
  {Olszewski}, {Mateo}, {Irwin}, \& {Battaglia}}]{deBoer14}
{de Boer}, T.~J.~L., {Tolstoy}, E., {Lemasle}, B., {Saha}, A., {Olszewski},
  E.~W., {Mateo}, M., {Irwin}, M.~J., \& {Battaglia}, G. 2014, \aap, 572, A10

\bibitem[{{de Boer} {et~al.}(2012){de Boer}, {Tolstoy}, {Hill}, {Saha},
  {Olsen}, {Starkenburg}, {Lemasle}, {Irwin}, \& {Battaglia}}]{deBoer12}
{de Boer}, T.~J.~L., {et~al.} 2012, \aap, 539, A103

\bibitem[{{de Marchi} {et~al.}(2006){de Marchi}, {de Angeli}, {Piotto},
  {Carraro}, \& {Davies}}]{deMarchi06}
{de Marchi}, F., {de Angeli}, F., {Piotto}, G., {Carraro}, G., \& {Davies},
  M.~B. 2006, \aap, 459, 489

\bibitem[{{Dotter} {et~al.}(2008){Dotter}, {Chaboyer}, {Jevremovi{\'c}},
  {Kostov}, {Baron}, \& {Ferguson}}]{dotter08}
{Dotter}, A., {Chaboyer}, B., {Jevremovi{\'c}}, D., {Kostov}, V., {Baron}, E.,
  \& {Ferguson}, J.~W. 2008, \apjs, 178, 89

\bibitem[{{El-Badry} {et~al.}(2015){El-Badry}, {Wetzel}, {Geha}, {Hopkins},
  {Kere{\v s}}, {Chan}, \& {Faucher-Gigu{\`e}re}}]{el-badry15}
{El-Badry}, K., {Wetzel}, A.~R., {Geha}, M., {Hopkins}, P.~F., {Kere{\v s}},
  D., {Chan}, T.~K., \& {Faucher-Gigu{\`e}re}, C.-A. 2015, ArXiv e-prints

\bibitem[{{Freeman}(2008)}]{freeman08}
{Freeman}, K.~C. 2008, The Messenger, 134, 28

\bibitem[{{Geha} {et~al.}(2013){Geha}, {Brown}, {Tumlinson}, {Kalirai},
  {Simon}, {Kirby}, {VandenBerg}, {Mu{\~n}oz}, {Avila}, {Guhathakurta}, \&
  {Ferguson}}]{geha13a}
{Geha}, M., {et~al.} 2013, \apj, 771, 29

\bibitem[{{Gosnell} {et~al.}(2014){Gosnell}, {Mathieu}, {Geller}, {Sills},
  {Leigh}, \& {Knigge}}]{gosnell14}
{Gosnell}, N.~M., {Mathieu}, R.~D., {Geller}, A.~M., {Sills}, A., {Leigh}, N.,
  \& {Knigge}, C. 2014, \apjl, 783, L8

\bibitem[{{Gratton} {et~al.}(2012){Gratton}, {Carretta}, \&
  {Bragaglia}}]{gratton12}
{Gratton}, R.~G., {Carretta}, E., \& {Bragaglia}, A. 2012, \aapr, 20, 50

\bibitem[{{Grebel}(1999)}]{grebel99a}
{Grebel}, E.~K. 1999, in IAU Symposium, Vol. 192, The Stellar Content of Local
  Group Galaxies, ed. {P.~Whitelock \& R.~Cannon}, 17--+

\bibitem[{{Helmi} {et~al.}(2006)}]{helmi06a}
{Helmi}, A., {et~al.} 2006, \apjl, 651, L121

\bibitem[{{Hurley-Keller} {et~al.}(1998){Hurley-Keller}, {Mateo}, \&
  {Nemec}}]{hurley-keller98}
{Hurley-Keller}, D., {Mateo}, M., \& {Nemec}, J. 1998, \aj, 115, 1840

\bibitem[{{Irwin} \& {Hatzidimitriou}(1995)}]{irwin95a}
{Irwin}, M., \& {Hatzidimitriou}, D. 1995, \mnras, 277, 1354

\bibitem[{{Karczmarek} {et~al.}(2015){Karczmarek}, {Pietrzy{\acute}ski},
  {Gieren}, {Suchomska}, {Konorski}, {G{\'o}rski}, {Pilecki}, {Graczyk}, \&
  {Wielg{\'o}rski}}]{karczmarek15}
{Karczmarek}, P., {et~al.} 2015, \aj, 150, 90

\bibitem[{Kauffmann {et~al.}(1993)Kauffmann, White, \&
  Guiderdoni}]{kauffmann93a}
Kauffmann, G., White, S. D.~M., \& Guiderdoni, B. 1993, \mnras, 264, 201

\bibitem[{{Kaviraj} {et~al.}(2007){Kaviraj}, {Kirkby}, {Silk}, \&
  {Sarzi}}]{kaviraj07}
{Kaviraj}, S., {Kirkby}, L.~A., {Silk}, J., \& {Sarzi}, M. 2007, \mnras, 382,
  960

\bibitem[{{Koch} {et~al.}(2006){Koch}, {Grebel}, {Wyse}, {Kleyna}, {Wilkinson},
  {Harbeck}, {Gilmore}, \& {Evans}}]{koch06}
{Koch}, A., {Grebel}, E.~K., {Wyse}, R.~F.~G., {Kleyna}, J.~T., {Wilkinson},
  M.~I., {Harbeck}, D.~R., {Gilmore}, G.~F., \& {Evans}, N.~W. 2006, \aj, 131,
  895

\bibitem[{{Kouwenhoven} {et~al.}(2009){Kouwenhoven}, {Brown}, {Goodwin},
  {Portegies Zwart}, \& {Kaper}}]{kouwenhoven09}
{Kouwenhoven}, M.~B.~N., {Brown}, A.~G.~A., {Goodwin}, S.~P., {Portegies
  Zwart}, S.~F., \& {Kaper}, L. 2009, \aap, 493, 979

\bibitem[{{Kroupa}(2001)}]{kroupa01}
{Kroupa}, P. 2001, \mnras, 322, 231

\bibitem[{{Larson}(2002)}]{larson02}
{Larson}, R.~B. 2002, \mnras, 332, 155

\bibitem[{{Leigh} {et~al.}(2013){Leigh}, {B{\"o}ker}, {Maccarone}, \&
  {Perets}}]{leigh13}
{Leigh}, N.~W.~C., {B{\"o}ker}, T., {Maccarone}, T.~J., \& {Perets}, H.~B.
  2013, \mnras, 429, 2997

\bibitem[{{Lemasle} {et~al.}(2012){Lemasle}, {Hill}, {Tolstoy}, {Venn},
  {Shetrone}, {Irwin}, {de Boer}, {Starkenburg}, \& {Salvadori}}]{lemasle12}
{Lemasle}, B., {et~al.} 2012, \aap, 538, A100

\bibitem[{{Mac Low} {et~al.}(1989){Mac Low}, {McCray}, \& {Norman}}]{maclow89}
{Mac Low}, M.-M., {McCray}, R., \& {Norman}, M.~L. 1989, \apj, 337, 141

\bibitem[{{Majewski} {et~al.}(2000){Majewski}, {Ostheimer}, {Patterson},
  {Kunkel}, {Johnston}, \& {Geisler}}]{majewski00b}
{Majewski}, S.~R., {Ostheimer}, J.~C., {Patterson}, R.~J., {Kunkel}, W.~E.,
  {Johnston}, K.~V., \& {Geisler}, D. 2000, \aj, 119, 760

\bibitem[{{Mateo} {et~al.}(1998){Mateo}, {Hurley-Keller}, \&
  {Nemec}}]{mateo98b}
{Mateo}, M., {Hurley-Keller}, D., \& {Nemec}, J. 1998, \aj, 115, 1856

\bibitem[{{Mateo}(1998)}]{mateo98a}
{Mateo}, M.~L. 1998, \araa, 36, 435

\bibitem[{{McConnachie}(2012)}]{mcconnachie12}
{McConnachie}, A.~W. 2012, \aj, 144, 4

\bibitem[{{McConnachie} \& {C{\^o}t{\'e}}(2010)}]{mcconnachie10}
{McConnachie}, A.~W., \& {C{\^o}t{\'e}}, P. 2010, \apjl, 722, L209

\bibitem[{{McMonigal} {et~al.}(2014){McMonigal}, {Bate}, {Lewis}, {Irwin},
  {Battaglia}, {Ibata}, {Martin}, {McConnachie}, {Guglielmo}, \&
  {Conn}}]{mcMonigal14}
{McMonigal}, B., {et~al.} 2014, \mnras, 444, 3139

\bibitem[{{Momany} {et~al.}(2007){Momany}, {Held}, {Saviane}, {Zaggia},
  {Rizzi}, \& {Gullieuszik}}]{momany07a}
{Momany}, Y., {Held}, E.~V., {Saviane}, I., {Zaggia}, S., {Rizzi}, L., \&
  {Gullieuszik}, M. 2007, \aap, 468, 973

\bibitem[{{Monelli} {et~al.}(2003){Monelli}, {Pulone}, {Corsi}, {Castellani},
  {Bono}, {Walker}, {Brocato}, {Buonanno}, {Caputo}, {Castellani}, {Dall'Ora},
  {Marconi}, {Nonino}, {Ripepi}, \& {Smith}}]{monelli03}
{Monelli}, M., {et~al.} 2003, \aj, 126, 218

\bibitem[{{Mu{\~n}oz} {et~al.}(2010){Mu{\~n}oz}, {Geha}, \&
  {Willman}}]{munoz10a}
{Mu{\~n}oz}, R.~R., {Geha}, M., \& {Willman}, B. 2010, \aj, 140, 138

\bibitem[{{Mu{\~n}oz} {et~al.}(2008){Mu{\~n}oz}, {Majewski}, \&
  {Johnston}}]{munoz08a}
{Mu{\~n}oz}, R.~R., {Majewski}, S.~R., \& {Johnston}, K.~V. 2008, \apj, 679,
  346

\bibitem[{{Mu{\~n}oz} {et~al.}(2006){Mu{\~n}oz}, {Majewski}, {Zaggia},
  {Kunkel}, {Frinchaboy}, {Nidever}, {Crnojevic}, {Patterson}, {Crane},
  {Johnston}, {Sohn}, {Bernstein}, \& {Shectman}}]{munoz06b}
{Mu{\~n}oz}, R.~R., {et~al.} 2006, \apj, 649, 201

\bibitem[{{Nonino} {et~al.}(1999){Nonino}, {Bertin}, {da Costa}, {Deul},
  {Erben}, {Olsen}, {Prandoni}, {Scodeggio}, {Wicenec}, {Wichmann}, {Benoist},
  {Freudling}, {Guarnieri}, {Hook}, {Hook}, {Mendez}, {Savaglio}, {Silva}, \&
  {Slijkhuis}}]{nonino99}
{Nonino}, M., {et~al.} 1999, \aaps, 137, 51

\bibitem[{{Pasetto} {et~al.}(2011){Pasetto}, {Grebel}, {Berczik}, {Chiosi}, \&
  {Spurzem}}]{pasetto11}
{Pasetto}, S., {Grebel}, E.~K., {Berczik}, P., {Chiosi}, C., \& {Spurzem}, R.
  2011, \aap, 525, A99

\bibitem[{{Piatek} {et~al.}(2003){Piatek}, {Pryor}, {Olszewski}, {Harris},
  {Mateo}, {Minniti}, \& {Tinney}}]{piatek03}
{Piatek}, S., {Pryor}, C., {Olszewski}, E.~W., {Harris}, H.~C., {Mateo}, M.,
  {Minniti}, D., \& {Tinney}, C.~G. 2003, \aj, 126, 2346

\bibitem[{{Pietrinferni} {et~al.}(2004){Pietrinferni}, {Cassisi}, {Salaris}, \&
  {Castelli}}]{pietrinferni04}
{Pietrinferni}, A., {Cassisi}, S., {Salaris}, M., \& {Castelli}, F. 2004, \apj,
  612, 168

\bibitem[{{Pilkington} \& {Gibson}(2012)}]{pilkington12}
{Pilkington}, K., \& {Gibson}, B.~K. 2012, in Nuclei in the Cosmos (NIC XII),
  227

\bibitem[{{Revaz} {et~al.}(2009){Revaz}, {Jablonka}, {Sawala}, {Hill},
  {Letarte}, {Irwin}, {Battaglia}, {Helmi}, {Shetrone}, {Tolstoy}, \&
  {Venn}}]{revaz09}
{Revaz}, Y., {et~al.} 2009, \aap, 501, 189

\bibitem[{{Rizzi} {et~al.}(2003){Rizzi}, {Held}, {Bertelli}, \&
  {Saviane}}]{rizzi03}
{Rizzi}, L., {Held}, E.~V., {Bertelli}, G., \& {Saviane}, I. 2003, \apjl, 589,
  L85

\bibitem[{{Saha} {et~al.}(1986){Saha}, {Monet}, \& {Seitzer}}]{saha86}
{Saha}, A., {Monet}, D.~G., \& {Seitzer}, P. 1986, \aj, 92, 302

\bibitem[{{Sales} {et~al.}(2010){Sales}, {Helmi}, \& {Battaglia}}]{sales10}
{Sales}, L.~V., {Helmi}, A., \& {Battaglia}, G. 2010, Advances in Astronomy,
  2010, 194345

\bibitem[{{Santana} {et~al.}(2013){Santana}, {Mu{\~n}oz}, {Geha},
  {C{\^o}t{\'e}}, {Stetson}, {Simon}, \& {Djorgovski}}]{santana13a}
{Santana}, F.~A., {Mu{\~n}oz}, R.~R., {Geha}, M., {C{\^o}t{\'e}}, P.,
  {Stetson}, P., {Simon}, J.~D., \& {Djorgovski}, S.~G. 2013, \apj, 774, 106

\bibitem[{Schlegel {et~al.}(1998)Schlegel, Finkbeiner, \& Davis}]{schlegel98a}
Schlegel, D.~J., Finkbeiner, D.~P., \& Davis, M. 1998, \apj, 500, 525

\bibitem[{{Smecker-Hane} {et~al.}(1996){Smecker-Hane}, {Stetson}, {Hesser}, \&
  {Vandenberg}}]{smecker-hane96}
{Smecker-Hane}, T.~A., {Stetson}, P.~B., {Hesser}, J.~E., \& {Vandenberg},
  D.~A. 1996, in Astronomical Society of the Pacific Conference Series,
  Vol.~98, From Stars to Galaxies: the Impact of Stellar Physics on Galaxy
  Evolution, ed. C.~{Leitherer}, U.~{Fritze-von-Alvensleben}, \& J.~{Huchra},
  328

\bibitem[{{Spada} {et~al.}(2013){Spada}, {Demarque}, {Kim}, \&
  {Sills}}]{spada13}
{Spada}, F., {Demarque}, P., {Kim}, Y.-C., \& {Sills}, A. 2013, \apj, 776, 87

\bibitem[{{Starkenburg} {et~al.}(2010){Starkenburg}, {Hill}, {Tolstoy},
  {Gonz{\'a}lez Hern{\'a}ndez}, {Irwin}, {Helmi}, {Battaglia}, {Jablonka},
  {Tafelmeyer}, {Shetrone}, {Venn}, \& {de Boer}}]{starkenburg10}
{Starkenburg}, E., {et~al.} 2010, \aap, 513, A34

\bibitem[{{Stetson}(1994)}]{stetson94a}
{Stetson}, P.~B. 1994, \pasp, 106, 250

\bibitem[{{Tolstoy} {et~al.}(2009){Tolstoy}, {Hill}, \& {Tosi}}]{tolstoy09}
{Tolstoy}, E., {Hill}, V., \& {Tosi}, M. 2009, \araa, 47, 371

\bibitem[{Unavane {et~al.}(1996)Unavane, Wyse, \& Gilmore}]{unavane96}
Unavane, M., Wyse, R. F.~G., \& Gilmore, G. 1996, \mnras, 278, 727

\bibitem[{{VandenBerg} {et~al.}(2015){VandenBerg}, {Stetson}, \&
  {Brown}}]{vandenberg15}
{VandenBerg}, D.~A., {Stetson}, P.~B., \& {Brown}, T.~M. 2015, \apj, 805, 103

\bibitem[{{Walker} {et~al.}(2009){Walker}, {Mateo}, {Olszewski},
  {Pe{\~n}arrubia}, {Wyn Evans}, \& {Gilmore}}]{walker09a}
{Walker}, M.~G., {Mateo}, M., {Olszewski}, E.~W., {Pe{\~n}arrubia}, J., {Wyn
  Evans}, N., \& {Gilmore}, G. 2009, \apj, 704, 1274

\bibitem[{{Weisz} {et~al.}(2014{\natexlab{a}}){Weisz}, {Dolphin}, {Skillman},
  {Holtzman}, {Gilbert}, {Dalcanton}, \& {Williams}}]{weisz14}
{Weisz}, D.~R., {Dolphin}, A.~E., {Skillman}, E.~D., {Holtzman}, J., {Gilbert},
  K.~M., {Dalcanton}, J.~J., \& {Williams}, B.~F. 2014{\natexlab{a}}, \apj,
  789, 147

\bibitem[{{Weisz} {et~al.}(2014{\natexlab{b}}){Weisz}, {Dolphin}, {Skillman},
  {Holtzman}, {Gilbert}, {Dalcanton}, \& {Williams}}]{weisz14b}
---. 2014{\natexlab{b}}, \apj, 789, 148

\bibitem[{{Xin} {et~al.}(2007){Xin}, {Deng}, \& {Han}}]{xin07}
{Xin}, Y., {Deng}, L., \& {Han}, Z.~W. 2007, \apj, 660, 319

\bibitem[{York {et~al.}(2000)}]{york00a}
York, D., {et~al.} 2000, \aj, 120, 1579

\bibitem[{{Zinn} \& {West}(1984)}]{zinn84}
{Zinn}, R., \& {West}, M.~J. 1984, \apjs, 55, 45

\end{thebibliography}
\end{document}